\documentclass{jfm}
\usepackage[usenames,dvipsnames,svgnames,table]{xcolor}
\usepackage{amsmath}
\usepackage{bbm}
\usepackage{rotating}
\providecommand{\boldsymbol}[1]{\mbox{\boldmath $#1$}}
\usepackage[normalem]{ulem} 
\usepackage{url}

\begin{document}

\newtheorem{lemma}{Lemma}
\newtheorem{corollary}{Corollary}
\renewcommand{\theequation}{\thesubsection.\arabic{equation}}
\newcommand{\com}[1]{\textcolor{green}{ #1}}
\newcommand{\req}[1]{\textcolor{red}{ #1}}

\newcommand{\MN}[1]{\textcolor{red}{ #1}}
\newcommand{\CD}[1]{\textcolor{blue}{ #1}}
\newcommand{\FG}[1]{\textcolor{orange}{ #1}}
\newcommand{\AL}[1]{\textcolor{olive}{ #1}}
\newcommand{\PT}[1]{\textcolor{violet}{ #1}}

\newcommand{\corr}[1]{\textcolor{blue}{ #1}}

\shorttitle{Fibers in confined channels} 
\shortauthor{Nagel et al.} 

\title{Oscillations of confined fibers transported in microchannels}

\author
 {M. Nagel \aff{1},
 P.-T. Brun \aff{1,2},
H. Berthet\aff{3},
A. Lindner\aff{3},
F. Gallaire \aff{1},
\and
C. Duprat\aff{4}
  }

\affiliation
{
\aff{1}
Laboratory of Fluid Mechanics and Instabilities, Ecole Polytechnique Federale de Lausanne, Lausanne 1015, Switzerland
\aff{2}
Department of Mathematics, Massachusetts Institute of Technology, Cambridge, Massachusetts 02139, USA
 \aff{3}
Physique et M\'ecanique des Milieux H\'et\'erog\`enes, UMR 7636, ESPCI Paris, PSL Research University, Universit\'e Paris Diderot, Universit\'e Pierre et Marie Curie, 10, rue Vauquelin, Paris, France
\aff{4}
Laboratoire d'Hydrodynamique (LadHyX), \'Ecole polytechnique, Palaiseau, France

}

\maketitle

\begin{abstract}
We investigate the trajectories of rigid fibers as they are transported in a pressure-driven flow, at low Reynolds number, in shallow Hele Shaw cells. The transverse confinement and the resulting viscous friction on these elongated objects, as well as the lateral confinement (i.e. the presence of lateral walls), lead to complex fibers trajectories that we characterize with a combination of microfluidic experiments and simulations using modified Brinkman equations. We show that the transported fiber behaves as an oscillator for which we obtain and analyze a complete state diagram.
\end{abstract}

\section{Introduction}
The motion of elongated particles (fibers) in viscous flows has been extensively studied, with examples ranging from the propulsion of microorganisms \citep{Lauga2009} to the clogging of arteries or stents with biofilm streamers \citep{Drescher2013}, the transport of fibers in fracture slits \citep{DAngelo2009}, the coupling of deformation and transport in flows \citep{Quennouz2015,lindnershelley2015} and the flow of dilute fiber suspensions in the paper-making industry \citep{Stockie1998}.
 
A prototypal situation is the sedimentation of a fiber in a Stokes flow. The fiber does not simply translate in the direction of gravity, but instead drifts at an angle depending on its orientation due to its drag anisotropy \citep{Cox1970}.  Another classical configuration, extensively studied since the pioneering work of \cite{Jeffery22}, is the rotation of a fiber in a 2D shear flow; the fiber has been shown to follow specific orbits, known as Jeffery orbits.
\begin{figure}
    \centerline{\includegraphics[width=0.5\textwidth]{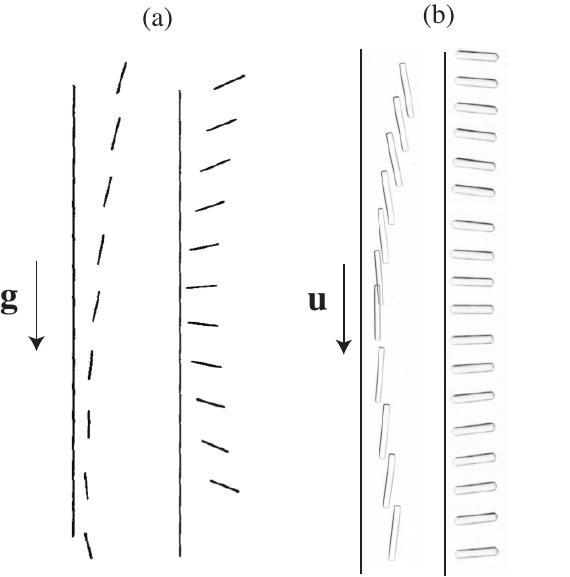}}
  \caption{ (a) Rotations of a fiber sedimenting near a wall, reprinted with permission from \cite{Russel77}. (b) Experimental chronophotographies of a fiber flowing in a microchannel exhibiting glancing and reversing motions near a wall.}
  \label{fig:illustration_russel}
\end{figure} 
Another situation, of particular interest for applications in microfluidics or porous media, is the motion of a fiber transported in a pressure-driven flow.
When the flow is unbounded, i.e. the fiber dimensions are small compared to the pore/channel size, the particle is simply advected at the speed of the imposed flow. In shallow Hele-Shaw cells or narrow pores, where the height of the fiber is comparable to the transverse channel height (so that the fiber nearly blocks the channel), the confinement causes viscous friction between the fiber and the surrounding walls. This friction reduces the velocity of the fiber, that is thus slower than the surrounding fluid, with a velocity that depends on its orientation: the fiber moves faster when oriented perpendicular to the flow direction than when parallel to the flow. This causes the fiber to drift when not aligned with the flow, with a drift direction opposite to that in sedimenting flows, where the particles move faster than the surrounding fluid \citep{Berthet2013}.
Similar observations were also made with other elongated objects, such as pairs of droplets \citep{Leman14} or rigid dumbbell particles \citep{Uspal2013}.

When transported in the flow, an object may also interact with lateral bounding walls (in opposition to the transverse confining walls). When sedimenting next to a wall, a fiber rotates away from the wall, in either a glancing or reversing motion depending on its initial inclination \citep{Russel77, DeMestre1975} as shown in Fig. \ref{fig:illustration_russel} (a). The analysis of these motions has been recently extended to the general case of oblate or prolate spheroids \citep{Mitchell2015}. In shear flows, a 'pole-vaulting' motion can be observed near the wall \citep{Stover90,Moses01}. In confined pressure-driven flows, elongated particles also rotate near walls, as was observed for fibers, dumbbell particles and pair of droplets \citep{helenethese, Uspal2013, Leman14}. As a consequence, an elongated object transported in a narrow channel oscillates between the channel walls. In particular, in our experiments we observe that the fibers drift and oscillate between opposite walls with motions resembling the glancing and reversing motions observed in sedimentation, as shown in Fig. \ref{fig:illustration_russel} (b).
However, a detailed inspection of these figures indicates noticeable differences; while in sedimentation, the angle of the fiber and the angle of its trajectory, though not equal, share the same sign, we observe in contrast that, in pressure-driven transport, the trajectory angle has an opposite sign with respect to the fiber orientation angle.

In this paper, we systematically investigate the transport of elongated fibers in a narrow channel, and study the effect of confinement on the fiber motion using a combination of experiments and simulations. Fiber trajectories are investigated in microfluidic experiments where fibers are fabricated in situ within microchannels with a photolithography process to ensure good control over the channel and particle properties. From a theoretical point of view, the $2D$~Stokes equations fail to describe the dynamics in this situation of strongly confined fibers. While the trajectory of a particle could be reproduced with $3D$~simulations, difficulties can arise due to the high mesh resolution needed to compute the velocity in the thin gap between the fiber and the channel walls. We thus propose a model combining the $2D$~Brinkman equations to a gap-flow model to take advantage of the robustness and numerical efficiency of a two dimensional approach while modeling the three dimensional effects in order to explore the role played by the key physical ingredients of the problem. Finally, we generate a complete state diagram of the fiber trajectories. 

We report several types of trajectories, especially glancing and reversing oscillations as presented in Fig. \ref{fig:illustration_russel}. In addition to these motions, we report new types of trajectories in the vicinity of the walls. In \textsection 2.1, we present our experimental setup; our theoretical formulation and numerical method are presented in \textsection 2.2. The validation of our model is given in \textsection 3. In \textsection 4, we describe our results, i.e. the different trajectories observed experimentally and numerically. Finally, our results are discussed in \textsection 5.

\section{Problem formulation and methods}
\begin{figure}
 \centerline{\includegraphics[width=\textwidth]{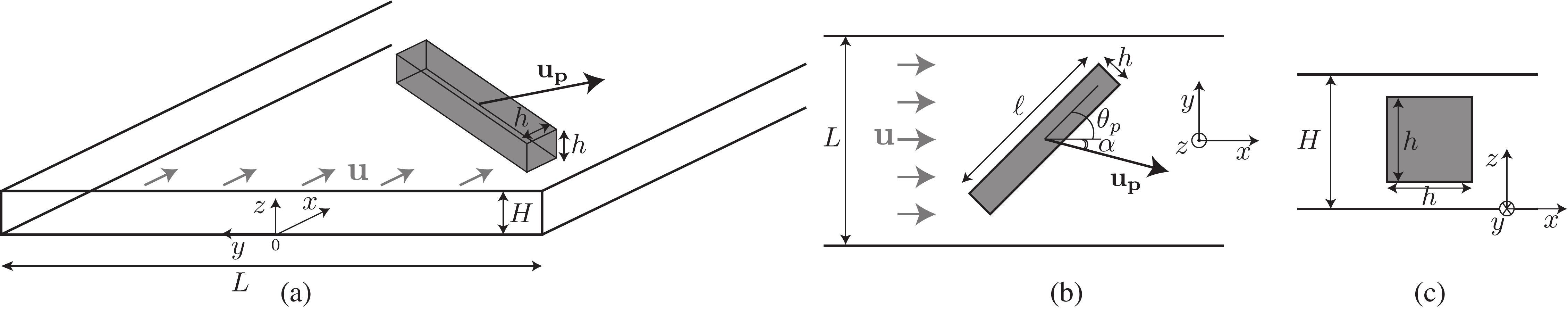}}
  \caption{(a) Sketch of a fiber in a microchannel, (b) top view and (c) cross-sectional view. \label{fig:not}}
\end{figure}

We study the transport of a fiber in a microchannel, as depicted in Fig. \ref{fig:not}, in the low Reynolds number limit. The channels are rectangular channels of height $H$ and width $L$ such that $H/L \ll 1$. We consider a fiber of length $\ell$ and width $h$ with a square cross-section, such that the aspect ratio $\ell/h$ is large. The fiber is transported by an externally imposed flow with mean flow velocity $\langle u \rangle$. The fiber is moving, at a speed ${\bf \hat u_p}$, in a direction given by the angle $\alpha$. Its orientation is given by the angle $\theta (t)$ and its trajectory by the position of its center of mass ($x_p (t)$, $y_p (t)$). The trajectories depend only on two geometrical parameters, the transversal confinement $\beta=h/H$ and the lateral confinement $\xi=\ell/L$. The gap size $bH$ is defined as the distance between the fiber and the top or bottom wall, i.e. $2b=1-\beta$. $\beta=1$ ($b=0$) corresponds to a particle filling the entire channel height and $\beta\simeq 0$ ($b=1/2$) to an infinitely thin object. Note that for $\xi\ll 1$ the effects of the lateral walls can be neglected at the center of the channel.

\subsection{Experimental set-up}

\begin{figure}
  \centerline{\includegraphics[width=0.7\textwidth]{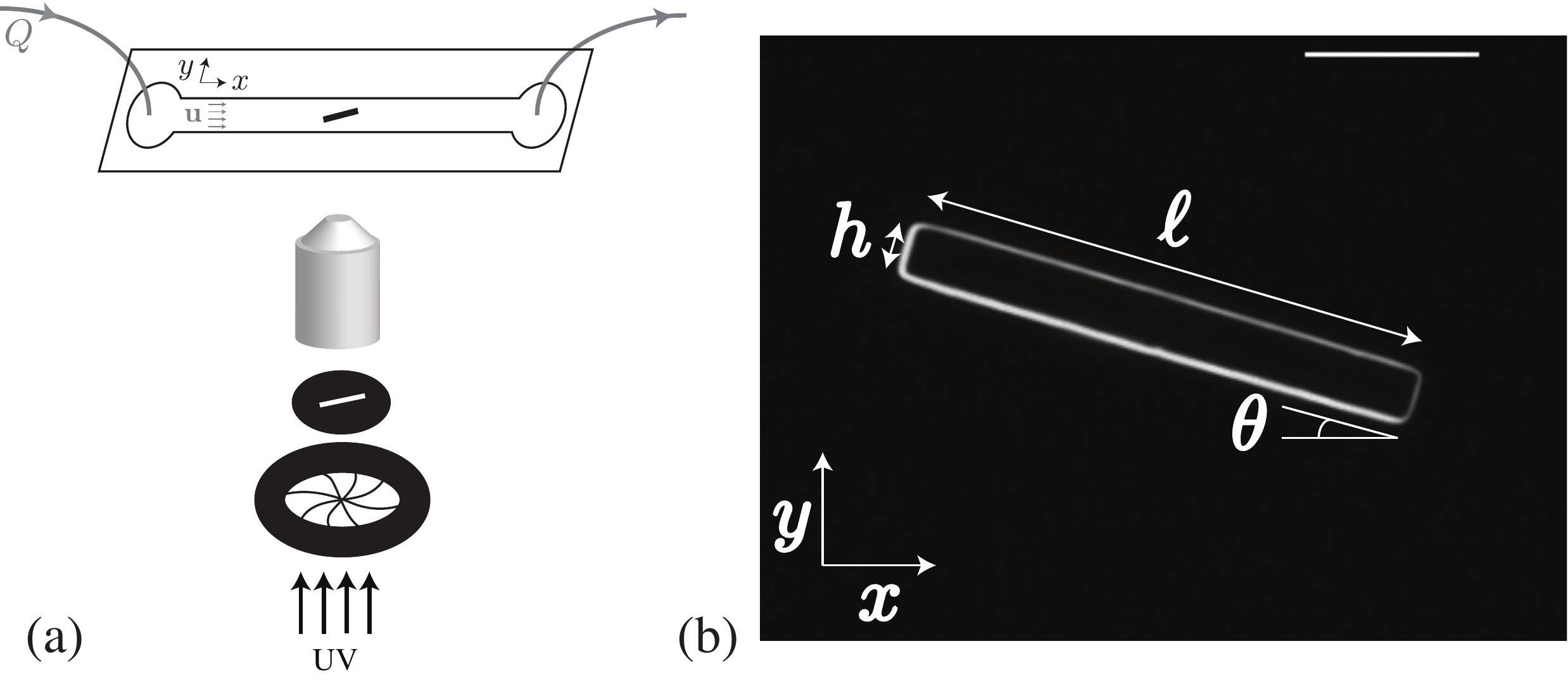}}
  \caption{(a) Experimental setup (b) Photography of a typical polymeric fiber using phase contrast microscopy (scale bar 200 $\mu$m). \label{fig:setup}}
\end{figure}

The microchannels are polydimethylsiloxane (PDMS) channels formed using molds fabricated with a micro-milling machine (Minitech Machinery), with an accuracy in channel height of $\pm 0.5 \,\mathrm{\mu m}$. The channels are bonded to a cover slide spin coated with a thin layer of PDMS in order to ensure identical boundary conditions on the four walls. We fabricate fibers of controlled geometry using the stop-flow microscope-based projection photolithography process developed by \cite{Dendukuri2007} (Figure \ref{fig:setup}(a)) and developped further by \cite{Berthet2016}. The channel is filled with a solution of oligomer and photo-initiator, and exposed to a pulse of UV light through a lithography mask placed in the field-stop position of the microscope. We use the method described in \cite{Duprat2015} to obtain rigid (i.e. high modulus) particles, with a solution of polyethylene glycol diacrylate (PEGDA, Aldrich) of average molecular weight 575 and 10$\%$ photo-initiator (Darocur 1173 (2-Hydroxy-2-Methylpropriophenone, Sigma)), exposed to UV light for over 500 ms with a Zeiss Axio Observer equipped with a UV light source (Lamp HBO 130W). We thus obtain a polymer fiber whose shape (length $\ell$ and width $h$) is determined by the shape of the mask (Figure \ref{fig:setup}(b)) with an accuracy within our optical accuracy $\pm 2 \,\mathrm{\mu m}$. We control the height of the fiber by taking advantage of the permeability of PDMS to oxygen that inhibits the polymerization, leaving a non-polymerized lubricating layer of constant thickness along the walls of the channel \citep{Dendukuri2008}. The height $h$, and the confinement $\beta$ are thus both determined by the height $H$ of the channel since the inhibition layer is of constant height $H-h=13\pm 1 \,\mu$m in our setup \citep{Berthet2013, Wexler2013,Duprat2015}. The fiber is thus fabricated at the center of the channel, i.e. the top and bottom lubricating layers have the same thickness. We estimate the error in $\beta$ to 0.05, which corresponds to a variation of the channel and/or fiber height of $\simeq 3~\mu$m. We adjust the shapes of the masks in order to ensure a square cross-section and an aspect ratio $\ell/h=8$ or $10$. The confinement $\xi$ is controlled by adjusting the width of the channel $L$. In all cases, we are in a Hele-Shaw configuration such that $H\ll L$. We vary the confinement by varying the channel height; all the other dimensions are changed accordingly to keep all aspect ratios constant. The resulting geometries are given in Table \ref{table:geo}.

\begin{table}
 \begin{center}
\def~{\hphantom{0}}
\begin{tabular}{cccccc}
~~$H ~(\pm 0.5)~ [\mu m]$~~ & ~~$L  ~(\pm 2)~ [\mu m]$~~ & ~~$h ~(\pm 2)~ [\mu m]$~~ & ~~$\ell  ~(\pm 2)~ [\mu m]$~~ & ~~$\beta ~(\pm 0.05)$~~ & ~~$\xi$~~\\[3pt]
38.5  & 1356  & 26  & 211  & 0.68  & 0.16  \\
51.0  & 1866  & 39  & 323  & 0.76  & 0.17  \\
51.0  & 2350  & 39  & 350  & 0.76  & 0.14  \\
60  & 460  & 47  & 470  & 0.78  & 1.02  \\
60  & 460  & 47  & 376  & 0.78  & 0.82  \\
60  & 600  & 47  & 376  & 0.78  & 0.63  \\
60  & 420  & 47  & 380  & 0.78  & 0.9  \\
61.5  & 2330  & 50  & 405  & 0.81  & 0.17  \\
68.5  & 2894  & 56  & 454  & 0.82  & 0.16  \\
70  & 700  & 57  & 456  & 0.81  & 0.65  \\
86.5  & 3873  & 77  & 616  & 0.89  & 0.16  \\
\end{tabular}
\caption{Geometries of the channels/fibers used in the experiments.}
\label{table:geo}
  \end{center}
\end{table}%

A precision pump (Nemesys, Cetoni) is used to drive the liquid in the channel. The fiber is fabricated at zero flow rate with a controlled initial position ($x_i,y_i$) and orientation $\theta_i$. When the flow is turned on, we monitor the trajectory ($x_p(t),y_p(t)$) and the angle $\theta(t)$. Distances are made dimensionless (dimensionless variables are denoted by a tilde) using $L/2$, such that $-1 \leq \tilde{y_p} \leq 1$.

\subsection{Theoretical formulation and simulations}

As in any fluid-solid interaction problem, the motion of the solid particle is bilaterally coupled to the flow velocity field, through the continuity of velocity and stresses at the fluid-solid interface. While the trajectory of a particle can be reproduced with $3D$~simulations of the Stokes equations, the mesh resolution requirements associated with the thin gap between the fiber and the channel walls can become prohibitive, in particular when $\beta$ approaches $1$. Alternatively, asymptotic approaches can be attempted (see \cite{Halpern91} for disk shape particles) but they need to take into account two small parameters $H/L \ll 1$ and $b \ll 1$. We thus propose a model combining the $2D$~Brinkman equations to a gap-flow model. The Brinkman equations, although not derived asymptotically from first principles, were shown, in the context of flows in thin channels around pancake-shaped droplets \citep{Boos97,Bush97}, to correctly capture the forces applied on the interface of the drop, providing a significant improvement with respect to the Darcy equations often used in Hele-Shaw cells \citep{Gallaire13}. The gap flow model discussed in section 2.2.4 combines a Couette and a Poiseuille flow and is reminiscent of the one used recently by \cite{Berthet2013}. In order to model the fiber trajectory, we first determine the resistance of a composite control volume that contains the fluid and the particle in the projected area of the particle. This modified resistance is then injected into the simulation of the liquid domain to compute the forces and the average displacement.
Our approach requires $b \ll 1$ in order to neglect the pressure-driven leakage flow compared to the mean flow in the microchannel. While this model cannot be rigorously derived from first principles, it will be shown a posteriori to be quantitatively valid in a rather large parameter range.   In the following, we first focus on the description of the fluid motion and stresses, then of the solid motion, before coupling them together.

\subsubsection{Depth-averaged flow equations}
\label{ch:brinkman}

We use the depth averaged $2D$~Brinkman equations to model the carrier flow in the fluid domain $\Omega_f=\Omega-\Omega_p$, where $\Omega_p$ is the particle's in-plane cross-section. This model assumes a parabolic velocity profile across the channel height,
\begin{equation}
\hat{\mathbf u}(x,y,z)={\mathbf u}(x,y) \frac{6z(z-H)}{H^2},\label{para}
\end{equation}
where ${\mathbf u}(x,y)$ is the depth-averaged in-plane velocity field. The $2D$~Brinkman equations are obtained by depth-averaging the $3D$~Stokes equations, assuming the aforementioned parabolic velocity profile (\ref{para}) across the height (so that the transverse $z$-velocity component is assumed negligible). This operation results in one term representing the viscous dissipation due to the Hele-Shaw confinement, as in Darcy's Law, and another in-plane viscous term similar to the one one found in the $2D$~Stokes equation \citep{Boos97,Bush97,Gallaire13}:

\refstepcounter{equation}\label{myeqn1}
\begin{equation}
  \mu \left( \nabla^2 \mathbf{u} -\frac{12}{H^2}\mathbf{u} \right) -\nabla p = 0, \quad \nabla \cdot \mathbf{u} = 0,
	 \tag{\theequation a,b} \label{eq:brman}
\end{equation}
where $\mu$ denotes the viscosity of the fluid.

Using a boundary integral approach, this equation can be reformulated as integral relations between the stresses and the velocities on the channel and particle boundaries ($\omega_c \cup \omega_p$) of the fluid domain $\Omega_f$, valid for any point ${\mathbf x}_0$ on these boundaries:
\begin{equation}
\oint_{\omega_p} \left(  {\mathbf T}_j{\mathbf n} \cdot {\mathbf u} - \sigma {\mathbf n} \cdot {\mathbf G}_j \right) ds +\oint_{\omega_c} \left(   {\mathbf T}_j{\mathbf n} \cdot {\mathbf u} - \sigma {\mathbf n} \cdot {\mathbf G}_j\right) ds= \frac{{\mathbf e}_j}{2} \cdot {\mathbf u}({\mathbf x}_0),
\label{BIM}
\end{equation}
where $j$ indicates the location $x_j,y_j$ of the Dirac delta of the Green's functions and ${\mathbf T}_j$, ${\mathbf G}_j$ are Brinkman's Green's functions for the stress and velocity \citep{Nagel2014}. We prescribe a known velocity field at the channel inlet, a parallel flow with constant normal stress at the outlet, and enforce a no slip condition on the channel walls. We therefore prescribe a known velocity/stress boundary value on the channel boundary $\omega_c$. 

\subsubsection{Rigid body motion of the particle and definition of a 'composite particle'}

There are  two important difficulties that arise when solving for the motion of rigid objects within the proposed depth-averaged approach: (i) the modeling and evaluation of the friction arising from the liquid films squeezed between the particle and the top and bottom walls of the channel  (they are prevalent in the system and therefore cannot be neglected) and (ii) connect the particle velocity and stresses to those of the depth-averaged flow. Consistent with our depth-averaged approach, we propose to apply a force and torque balance on a {\it composite control volume}, i.e. a slice which includes the rigid particle and the fluid in the thin films. This slice ranges from $[0,H]$  with the intervals $[0,bH]$ and $[(1-b)H,H]$ occupied by the liquid phase and the interval $[bH,(1-b)H]$ occupied by the rigid particle. With this model we implicitly assume that the particle finds its equilibrium position in the center of the channel so that the center plane of the Hele-Shaw cell $z=0$ is also a symmetry plane of the particle. This assumption follows from neglecting gravity effects (small Galilei number) and supposing that the particle will maintain least dissipation by staying centered.

The $2D$ representation of the shape of the composite particle is given by its surface $\Omega_p$ of area $A_p$ with boundary $\omega_p$. The depth-averaged velocity of any point $(x,y) \in  \Omega_p$ is denoted ${\mathbf u}_p$ and is representative of both the particle and the liquid layers enclosed between the particle and the walls. As a rigid body in planar motion our composite particle has three degrees of freedom: the velocities $U_p, V_p$ in $x$ and $y$ direction and the angular velocity around its center $\dot \theta_p$.
Writing the velocity field in the frame of its barycenter yields
\begin{equation}
	\mathbf{u}_p(x,y) = \left( \begin{array}{ccc}
		1 & 0 & - y \\ 0 & 1 & x
	\end{array} \right) \left( \begin{array}{c}
	U_p \\ V_p \\ \dot \theta_p
	\end{array} \right).
	\label{eq:avbodyvel}
\end{equation}
It is important to recall that  $\mathbf{u}_p(x,y)$ is not the velocity of a material point of the particle $(x,y,z)$ (for $z \in [bH,(1-b)H]$), but rather the depth-averaged velocity of the vertical slice containing $(x,y)$, i.e. the depth-averaged velocity of a slice of the composite particle. Similarly, $U_p$ and $V_p$ are the rigid body velocities of the composite particle, not of the particle itself, as later detailed in section \ref{velo}. Applying the kinematic boundary conditions at the composite-particle/fluid boundary forbids any leakage flow in the thin gaps when the particle is at rest. This strong hypothesis is reasonable in the thin gap regime ($b \ll 1$), since the hydraulic resistance trough the thin gaps is $\mathcal{O}(b^{-2})$ larger than that around the particle.

\subsubsection{Stress on the composite particle}

The depth-averaged stress density per unit length $\mathbf{f}$ that the fluid exerts on the composite particle lateral walls $\omega_p$  results from viscous and pressure stresses. We write  $\mathbf{f} = \sigma \mathbf{n}$, with the stress tensor $\sigma=-p \mathbbm{I} + 2\mu \mathbbm{D}$, where $\mathbbm{D}$ is the symmetric part of the depth-averaged rate-of-strain tensor, and $\mathbf{n}$ the normal on the rectangle $\omega_p$. The total depth-averaged force and momentum acting on the composite particle are obtained by integrating along the rectangle perimeter $\omega_p$, parametrized by its local abscissa~$s$
\begin{eqnarray}
	\mathbf{F}_\parallel & = & H \oint \limits_{\omega_p} \sigma \mathbf{n} ds = H \oint \limits_{\omega_p}  \left( \begin{array}{c}
		f_x  \\
		f_y
	\end{array} \right) ds,
	\label{eq:forcepara}	
	 \\
	M_\parallel & = & H \oint \limits_{\omega_p} \sigma  \mathbf{n} \times \mathbf{x} ds = H \oint \limits_{\omega_p} -f_x\, y+f_y \, x \,ds.
	\label{eq:couplepara}
\end{eqnarray}
These force and momentum components are only a part of the total force as they only incorporate the depth-averaged force on the composite particle side faces. The total force balance requires to also include the force and momentum exerted by the channel walls touching the top and bottom composite particle faces.

When deriving the depth-averaged fluid equations \eqref{eq:brman} away from the particle, we assumed a parabolic velocity profile in the $z$ direction. Similarly, we now use an Ansatz velocity profile $q(z)$ for the flow in the gap between the wall and the object. We leave $q(z)$ unspecified at this stage, and only require that its mean value over the channel height equals $1$, defining the full $3D$ velocity $\hat  {\bf u}_p(x,y,z) =  {\bf u}_p(x,y) \, q(z)$. We further denote the gradient of $q(z)$ at the bottom wall as $q^\prime = d q(0)/d z$. The forces exerted by the top and bottom walls onto the composite particle are thus
\begin{eqnarray}
	F_{\perp,x} &=& 2 \mu \int \limits_{\Omega_p} \frac{\partial \hat u}{\partial z} \Big|_{z= 0} \; dA = 2 \mu \int \limits_{\Omega_p} (U_p - y \dot \theta_p ) q^\prime \; dA = 2 \mu U_p \, q^\prime \, A_p,
\label{eq:fiber_forcex}
\\
	F_{\perp,y} &=&  2 \mu \int \limits_{\Omega_p} \frac{\partial \hat v}{\partial z} \Big|_{z=0} \; dA = 2 \mu \int_{\Omega_p} (V_p+x \dot \theta_p) q^\prime \; dA =  2 \mu V_p \, q^\prime \, A_p,
	\label{eq:fiber_forcey}
\end{eqnarray}
where for symmetry reasons the total force is twice the force exerted on the bottom wall. Note that these expressions depend only on the depth-averaged velocities $U_p$ and $V_p$ of the composite particle, the gradient $q^\prime$ and the area that faces the top or bottom wall $A_p$. Because the particle rotates around its center, the domain integrals that depend on $\dot \theta$ are in effect equal to  zero. 

Similarly, we derive the torque
\begin{equation}
	M_\perp = 2 \mu \int \limits_{\Omega_p} \frac{\partial (\hat v \,x -\hat u \,y)}{\partial z} \Big|_{z= 0} \; dA = 2 \mu \dot \theta_p q^\prime  \int \limits_{\Omega_p} (x^2+y^2) \; dA =2 \mu \, \dot \theta_p \, q^\prime \, T_p,
	\label{eq:fiber_couple}
\end{equation}
where $T_p$ is a second order moment $T_p = \int \limits_{\Omega_p} (x^2+y^2) \; dA$, which can also be obtained by integration on the domain boundary $T_p = \oint \limits_{\omega_p} \left( x y^2 \, n_x +x^2 y\,n_y  \right) ds$.

In the vanishing Reynolds number limit that we consider here, the composite particle has to be force free and torque free. These conditions translate in a set of 3 equations (equations 
\eqref{eq:fiber_forcex}, \eqref{eq:fiber_forcey} and \eqref{eq:fiber_couple})
for the unknowns $U_p, V_p, \dot{\theta}, \left. {\mathbf f}\right|_{\omega_p}$. 

The averaged velocities on the particle interface are the ones introduced in eq.\eqref{eq:avbodyvel}.
Finally, the profile $q(z)$ will be determined in the next paragraph.


\subsubsection{Gap flow model}
\label{ch:gapflow}

We need to determine the velocity profile $q(z)$ on order to close the system: its value will determine the shear at the wall $q^\prime$ but also the ratio between the velocity of the composite particle ${\mathbf U}_p$ and the particle velocity $\hat{\mathbf U}_p$. This derivation will be made using an Ansatz for $q(z)$ .

First, as it is the simplest non-trivial profile, we assume that $q(z)$ is a linear velocity profile that ensures the compatibility condition $\frac{1}{H}\int_{0}^{H} q_l(z)dz=1$, given by
\begin{equation}
q_{\, \text{linear}}(z)= \frac{4H}{H^2-h^2} \left( z \mathbbm{1}_{[0,bH]} + (H-z) \mathbbm{1}_{[(1-b)H,H]} \right)+ \frac{2H}{H+h} \mathbbm{1}_{[bH,(1-b)H]},   \label{broken}
\end{equation}
where $\mathbbm{1}$ is the indicator function, a function that takes the value $1$ when $z$ is within the limits $[a,b]$ and $0$ elsewhere. This profile imposes a Couette flow in the liquid films and rigid motion in the particle. This linear velocity profile in the gap is reasonable 
for high confinement values ($\beta \sim 1$). In particular, one retrieves $q(H/2) = 1$ as expected for $\beta = 1$ and $b=0$ (an object that fills the entire channel height).
However, this Ansatz yields an underestimation of  the dissipation in the limit of small values of $\beta$.
Indeed, one expects $q(H/2) = 3/2$ for $\beta=0$ and $b=1/2$, an infinitely thin object, while the broken-line profile \ref{broken} gives $q(H/2)=2$.

As an improvement to our model we add a Poiseuille profile to our Ansatz, in the spirit of the gap model of \cite{Berthet2013}. This parabolic profile is first of undetermined amplitude,then truncated at distance $b\,H$ from the top or bottom wall and finally rescaled in order to ensure that the average of $q(z)$ over the channel height is $1$, such that
\begin{equation}
	q(z) = C \left[\left(1 - \frac{z}{H} \right) \frac{z}{H} \left(\mathbbm{1}_ {[0,bH]} +
	 \mathbbm{1}_{[(1-b)H,H]}\right)+ (1 - b)b\mathbbm{1}_{[bH,(1-b)H]}\right],
	\label{eq:qprofile}
\end{equation}
where the constant $C$ that ensures $\frac{1}{H}\int_{0}^{H} q_l(z)dz=1$ is given by
\begin{equation}
C = \frac{1}{b-2b^2+4/3 b^3}=\frac{6}{1-\beta^3}.
\label{eq:detC}
\end{equation}
\begin{figure}
		\centering
		\includegraphics{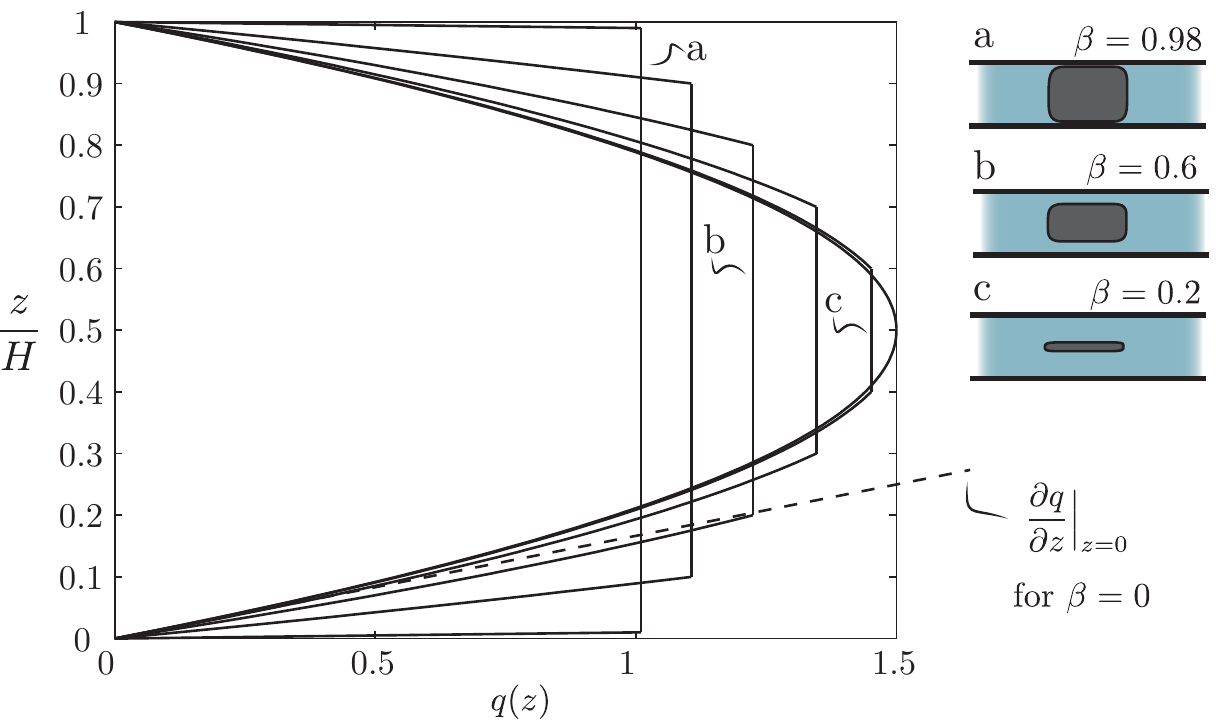}
		\caption{Model velocity profiles in the shallow direction in the presence of a rigid object. The flat region is where the object is located and hence the velocity is constant. Profiles are shown for $\beta=0.98, 0.8, 0.6, 0.4, 0.2$ and $0$. For $\beta=0$ the object is infinitely flat and the velocity profile becomes a parabola. The velocity gradient for $\beta = 0$ is illustrated by a tangent $- - -$ line.}
 	\label{fig:fiber-vel-prof}
\end{figure}
The velocity gradient at the wall is then given by
\begin{equation}
	q^\prime = \frac{6}{H(1-\beta^3)},
	\label{eq:fprime}
\end{equation}
which should be plugged into equations \eqref{eq:fiber_forcex}-\eqref{eq:fiber_couple} to close for the gap flow model. Figure \ref{fig:fiber-vel-prof} shows the obtained velocity profiles for different confinements. The profiles exhibit a smooth transition from a trapezoidal profile for the strongly confined particle to a parabola for the unconfined particle. For increasing confinements $\beta$ the ratio of particle to mean velocity $q(H/2)$ decreases, while the velocity gradient at the wall increases.

In summary, the velocity field in the entire domain takes the following form
\begin{eqnarray}
\hat{\mathbf u}(x,y,z)={\mathbf u}(x,y) \frac{6}{H^2} z(z-H) \mathbbm{1}_{[0,H]\times\Omega_f}+
{\mathbf u}_p(x,y) \frac{3}{2}\frac{1-\beta^2}{1-\beta^3} \mathbbm{1}_{[bH,(1-b)H]\times\Omega_p}\\\nonumber
+{\mathbf u}_p(x,y) \frac{6}{1-\beta^3}\left(1 - \frac{z}{H} \right) \frac{z}{H} \left(\mathbbm{1}_ {[0,bH]\times \Omega_p} + \mathbbm{1}_{[(1-b)H,H]\times \Omega_p} \right).
\end{eqnarray}

\subsubsection{Rigid particle velocity}
\label{velo}

We now need to derive the velocity $\hat{\mathbf{U}}_p$ and the rotation velocity $\hat{\dot{\theta}}_p$ of the rigid particle from the velocity $\mathbf{U}_p$ and rotation rate $\dot{\theta}_p$ of the composite particle. The particle velocities are deduced from the ratio of the velocity of the points $(x,y) \in \Omega_p$ in the particle $\hat{\mathbf{u}}_p (x,y,z\in[bH,(1-b)H])$ to the depth-averaged velocity $\mathbf{u}_p (x,y)$, that is derived from the mean velocity and the confinement $\beta$ using eq.\eqref{eq:detC} such that
\begin{equation}
	\hat{\mathbf{u}}_p(x,y)= \mathbf{u}_p (x,y) \frac{3}{2} \frac{1+\beta}{1+\beta+\beta^2}.
		\label{eq:uprime}
\end{equation}	
The rigid body velocities and rotation of the particle are deduced from the averaged velocity of the composite particle according to
\begin{equation}
	(\hat{U}_p,\hat{V}_p,\hat{\dot{\theta}}_p)= \frac{3}{2} \frac{1+\beta}{1+\beta+\beta^2} (U_p,V_p,\dot{\theta}_p).
		\label{eq:uprimeis}
\end{equation}	
Since the particle moves at a different velocity than the average flow velocity, we use these velocities, designated by a hat, to deduce the particle motion and update its location. 

\subsubsection{Numerical simulations}
For the numerical resolution of the differential equation, we propose the boundary element method (BEM) as this technique is well suited to problems with evolving interfaces. The BEM makes use of the Green functions of the Brinkman equation and has proven successful to simulate droplets in shallow microchannels \citep{Nagel2014}.

The equations of the problem are non-dimensionalized with the inflow velocity $u_\infty$, the characteristic length of the channel $L/2$ and viscosity $\mu$, such that equations \eqref{eq:brman} read
\refstepcounter{equation}\label{myeqn2}
\[
	\left( \nabla^2  \mathbf{\tilde u} -k^2 \mathbf{\tilde u} \right) -\nabla \tilde p = 0,
	 \quad \nabla \cdot \mathbf{\tilde u} = 0.
	 \tag{\theequation a,b} \label{eq:ndbrman}
\]
where  $k= \frac{\sqrt{12}}{\tilde h}$.
Non-dimensional variables are denoted by a tilde. For instance, the velocity is non-dimensionalized by $\mathbf{\tilde u} = \mathbf{u}/u_\infty $ and the channel height by $\tilde h = 2H/L = 2 \ell/L \cdot H/ \ell \cdot h/ \ell = 2 h/ \ell \cdot \xi /\beta$.

The computational domain extends from $\tilde x=-5$ to $5$ with $1500$ elements per side wall and from $\tilde y=-1$ to $1$ with $200$ elements per inflow or outflow. The fiber itself is discretized with $800$ elements. Its position $y_p$ and orientation $\theta_p$ evolve over time. Its displacement in the  $x$-direction is recorded but artificially cancelled out numerically owing to the invariance of the problem with respect to $x$. This is advantageous numerically and in particular allows the fiber to remain centered in the computational domain. The non-dimensional time step is varied from $0.001$ for fibers that approach the wall very closely to $0.05$ for fibers that are at one fiber diameter away from the wall and more. Depending on the situation, a one step Euler explicit scheme or a two-step scheme, Heun's Rule, are used. We observe that for Euler's scheme the amplitudes in $y_p$ and $\theta_p$ show a slight increase over time, whereas Heun's Rule rather shows a slight decrease.


Note that in the following we work exclusively with nondimensional variables and omit the tilde. 

\section{Validation of the model}

\subsection{Fiber Velocity}

In order to validate our model, we focus on the advection of a fiber oriented parallel ($\theta=0^{\circ}$) and perpendicular ($\theta=90^{\circ}$) to the flow direction, in an infinitely wide channel and in the presence of lateral walls.

\begin{figure}
\centering
	\includegraphics[width=0.8\textwidth]{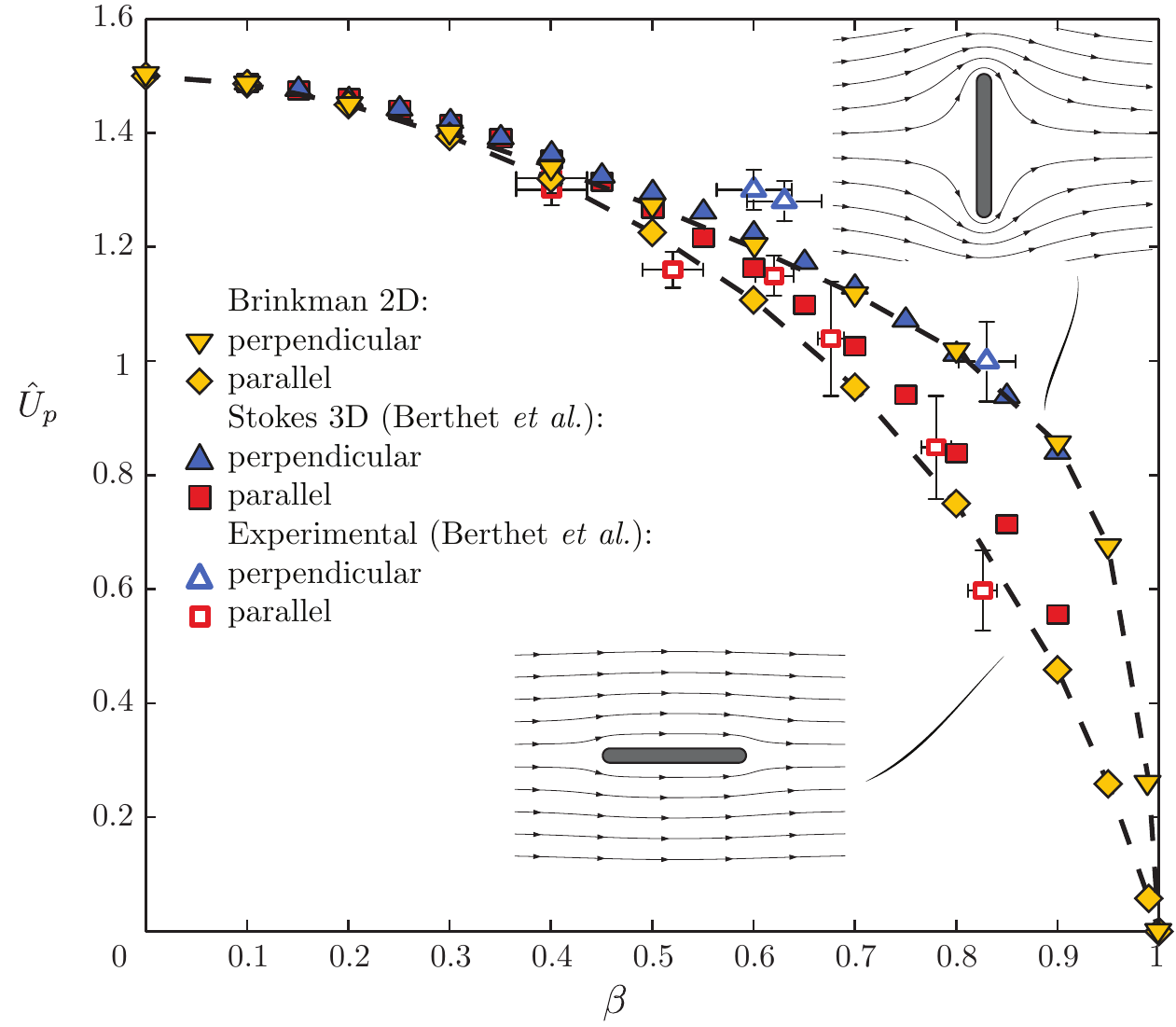}
	\caption{Fiber velocities for varying confinement $\beta$ in an infinitely large channel, computed with Brinkman $2D$ (yellow symbols) and compared to $3D$~Stokes simulations and experiments \citep{Berthet2013}. The fiber is either parallel or perpendicular to the flow.}
		\label{fig:fibervel}
\end{figure}

In an infinitely wide channel, the fiber does not change orientation and simply translates with a velocity that depends on the confinement $\beta$, as was obtained experimentally and numerically (3D Stokes simulations) by \cite{Berthet2013} (Figure \ref{fig:fibervel}). When the confinement is large, i.e. $\beta\geq0.6$, fibers perpendicular to the flow are transported at a velocity higher than those transported parallel to the flow. This transport anisotropy varies with the confinement $\beta$ and is an essential ingredient for the fiber dynamics described in the following.
Our model is in good agreement with the experimental results as well as the $3D$~Stokes simulations from \cite{Berthet2013}, thereby validating the assumptions made when deriving the model. 

\begin{figure}
		\centering
		\includegraphics[width=\textwidth]{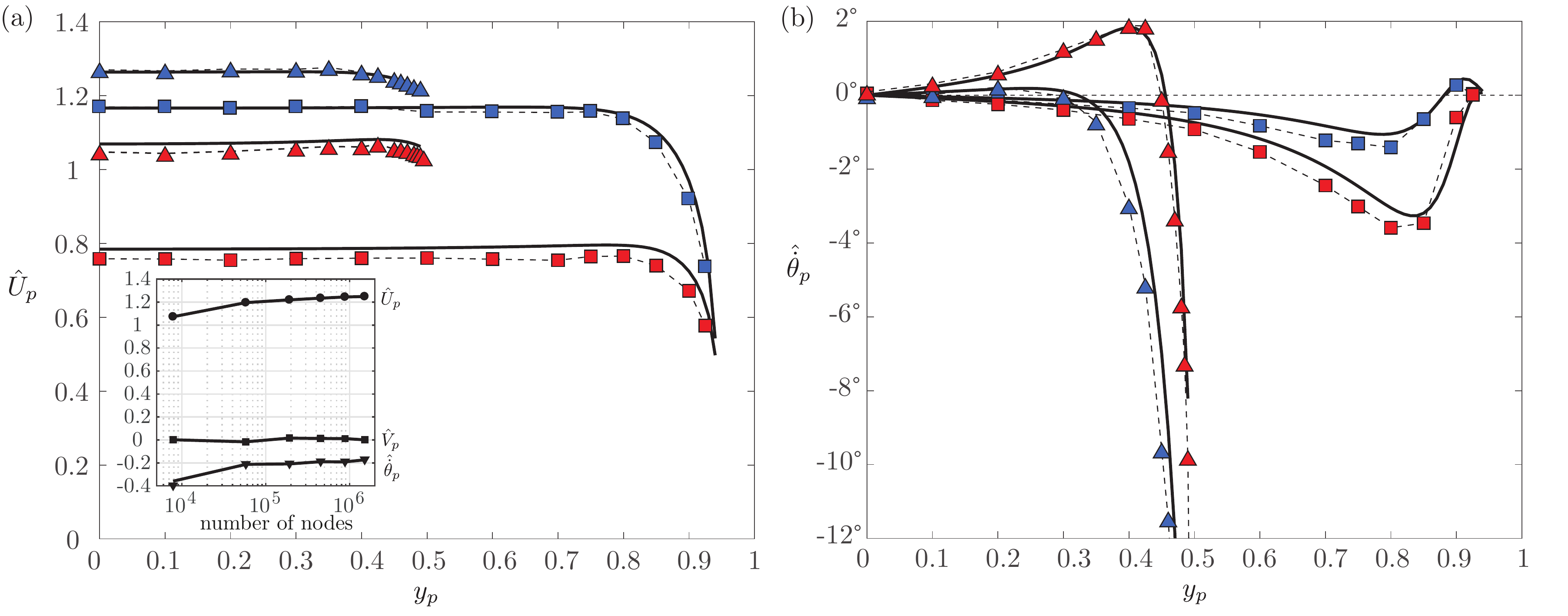}
		\caption{Comparison between streamwise fiber velocity and rotation rate in a channel with lateral walls calculated with the depth-averaged model (plain line) and a full 3D calculation (symbols). The aspect ratio $\xi=1/2$ for all four cases, $\bigtriangleup$ indicates fibers perpendicular and $\square$ parallel to the flow. Blue filled symbols stand for $\beta=0.6$ and red filled symbols for $\beta=0.8$. The inset shows the convergence for a perpendicular fiber with $\xi=0.5, \beta=0.8$ and $y_p=0.45$.}
		\label{fig:comp3d}
	\end{figure}

In order to validate our model in the presence of lateral walls, that may affect the fiber streamwise velocity as well as induce a rotation (as presented in Fig.\ref{fig:illustration_russel}), we compare our results to full 3D Finite Element Method (FEM) calculations. The FEM calculations are described in the appendix \ref{ap:fem}. We compute the advection velocity $\hat{U}_p$, as well as the rotation rate $\hat{\dot{\theta}}_p$ as a function of the position $y_p$. The results are shown in figure \ref{fig:comp3d}. The lateral confinement is fixed to $\xi=1/2$ and two transversal confinements $\beta=0.6$ and $0.8$ are shown.

The streamwise velocities obtained with the 3D simulation of the Stokes equation and with the Brinkman model are in excellent agreement (Fig. \ref{fig:comp3d}(a)); the agreement between both calculations is within a few percents, even near the wall. The rate of rotation also shows a good agreement even though the relative errors are more pronounced (Fig. \ref{fig:comp3d}(b)).

\subsection{Validity of the model}


As explained in section  \ref{ch:brinkman} and \ref{ch:gapflow}, the proposed depth-averaged model contains two main ingredients,
a Brinkman approximation for the suspending fluid and a gap-flow model for the flow in the thin gaps. Both are matched through a composite control volume description of the particle, on which the balance of forces is applied. Comparisons with prohibitive memory, time and energy-consuming  3D calculations show its quantitative prediction capacity, even when the condition $b \ll 1$ is violated.

One of the main interests of the Brinkman approximation is indeed that it correctly captures the dominant forces, when pushed out of its domain of validity, i.e. on the particle boundary.  Thanks to the relative balance of the Laplacian term with respect to the Darcy term proportional to $k^2$, the Brinkman equations  emulate the boundary layer near the particle boundary, with correct physical scalings. While in the bulk, the Laplacian term is negligible with respect to the Darcy contribution, it becomes significant in the vicinity of boundaries.  Thereby, 
it accounts for tangential stresses, which, in turn, enable the computation of the fiber velocity. In stark contrast, a Darcy like model would fail. Even in the simplest case scenario, when the fiber is aligned parallel to the flow direction, a Darcy like approach would be inaccurate. Such an approach would only account for the pressure acting on the tip regions of the fiber and therefore underestimate the fiber velocity.

The thin-gap model which is used in our gap flow model and therefore for the composite particle combines a Couette flow and a Poiseuille contribution which is designed to ensure continuity with the large gap limit $\beta=0$.

For dynamical simulations of a moving fiber in a channel, the proposed depth-averaged method uses relatively few unknowns located at the boundaries, which makes it significantly faster than a $3D$~Stokes simulation. This feature is of paramount importance as we aim to explore a large parameter space to develop a physical description of the system.


\section{Results}
\subsection{Fiber drift}
\begin{figure}
\centering
	\includegraphics[width=0.5\textwidth]{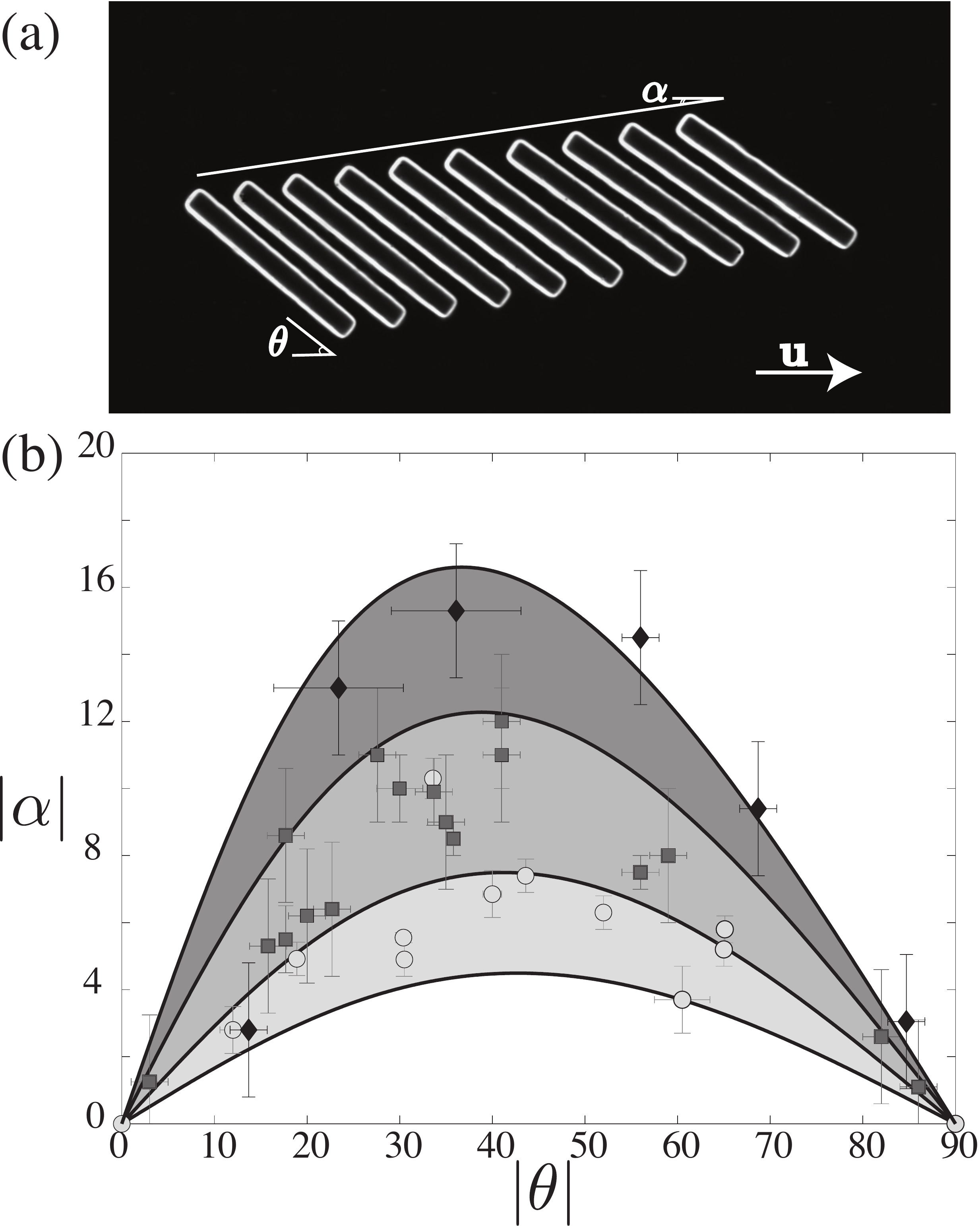}
	\caption{(a) Chronophotography of a fiber drifting. (b) Drift angle as a function of the fiber orientation for $\beta=0.75 \pm 0.05$ (light gray circles), $\beta=0.8 \pm 0.05$ (dark gray squares) and $\beta=0.86 \pm 0.05$ (black diamonds). The solid curves are given by eq.\eqref{eq:angledrift} with $B(\beta=0.7)=1.17$, $B(\beta=0.78)=1.3$, $B(\beta=0.85)=1.54$ and $B(\beta=0.89)=1.8$ given by our Brinkman model where $B=u_{\perp}/u_{\parallel}$. The shaded regions thus correspond to $0.7<\beta<0.78$ (light gray), $0.78<\beta<0.85$ (medium gray) and $0.85<\beta<0.89$ (dark gray).}
		\label{fig:drift}
\end{figure}

We first consider fibers transported in wide channels (with lateral confinement $\xi = O(10^{-1})$) far from the walls. For $\theta=0^{\circ}$ or $90^{\circ}$, the fiber moves along the $x$ axis only. For other angles, the fiber is advected downstream, but also has a vertical motion; the fiber drifts towards the walls of the channel (Figure \ref{fig:drift} (a)). The orientation angle $\theta$ can not be locked in experiments, and the fiber is subjected to small perturbations; we thus focus on trajectories where the orientation angle $\theta$, and thus the drift angle $\alpha$, remain constant for a significant travelled distance.

The evolution of the drift angle with the fiber orientation $\theta$ and the confinement $\beta$ is given in Fig. \ref{fig:drift} (b). We note that as the confinement increases, the drift angle $\alpha$ increases for a given orientation $\theta$. We find that the orientation at which the drift is maximum $\theta_{max}$ varies with the confinement but remains close to 45$^{\circ}$.

The drift finds its source in that a fiber is transported faster with its axis perpendicular to the flow than with its axis aligned with the flow. As shown in Fig. \ref{fig:fibervel},  we find that the ratio  $B(\beta)= u_{\perp}/u_{\parallel} >1$ . To evaluate the drift angle we recast the components of the fiber velocity in the lab referential, along $\bf{e}_x$ and $\bf{e}_y$, 
\begin{eqnarray}
u_{px}&=&u_{\parallel}\cos^2\theta+u_{\perp}\sin^2\theta\\
u_{py}&=&(u_{\parallel}-u_{\perp})\sin\theta\cos\theta,
\label{eq:udrift}
\end{eqnarray}
and find the drift angle
\begin{equation}
\label{eq:angledrift}
\tan\alpha=\frac{u_{py}}{u_{px}}=\frac{(1-B)\cos\theta\sin\theta}{\cos^2\theta+B\sin^2\theta}.
\end{equation}
The evolution of $\theta_{max}$ is governed by the equation:
\begin{equation}
\theta_{max}=\pm 2 \arctan\left[\left( 1+2B-2(B(1+B))^{1/2}\right)^{1/2}\right].
\end{equation}
We compute the value of $B$ using our $2D$~Brinkman model (Fig. \ref{fig:fibervel}) and find that our theoretical predictions for $\alpha$ are in fair agreement with experimental values (Fig~\ref{fig:drift}). 
For low transversal confinement ($\beta \leq 0.6$), parallel and perpendicular velocities are close, i.e. $B\simeq 1$ (Figure \ref{fig:fibervel}), so that the magnitude of the drift angle remains small. The drift angle then strongly increases with increasing confinement.
Therefore, a small variation in $\beta$ (i.e. of order of our  accuracy of 0.05) leads to high variations of $\alpha$, as indicated by the shaded regions in Fig.~\ref{fig:drift} , which also explains the scatter in the experimental data. Changing the confinement allows us to tune the drag anisotropy, and thus the magnitude of the drift.

\subsection{Effect of the bounding walls}
The presence of lateral walls modifies the flow field, and thus affects the fiber trajectory, inducing in particular a rotation of the particle. We first focus on the trajectories near the center of the channels, then describe the behaviour of the fibers when placed in the vicinity of the walls.

\label{sec:pendulum}
\subsubsection{Oscillations around $\theta=0^\circ$}

\begin{sidewaysfigure}
		\centering
		\vspace{5.15in}
		\includegraphics[width=\textwidth]{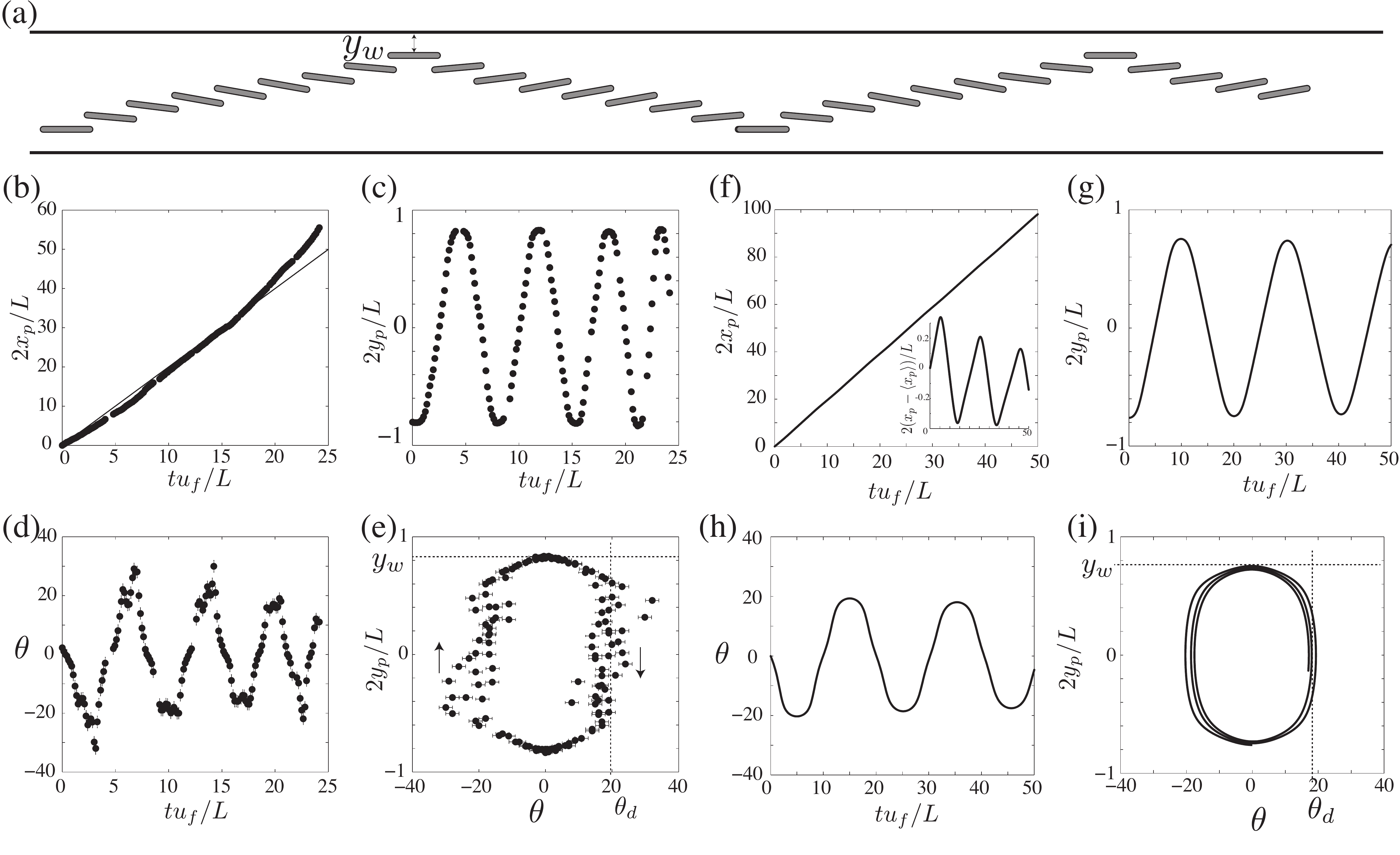}
		\caption{(a) Sketch of the oscillation mode. (b-i) Trajectory of a fiber advected by a mean flow for $\beta=0.81$, $\xi=0.81$ and $\theta_i=22^{\circ}$ obtained experimentally (b-e) and numerically (f-i): (b) axial position $x_p(t)$. (f) Deviation around the mean position $x_p-\langle x_p \rangle$ (position $x_p(t)$ given in inset). (c,g) streamwise position $y_p(t)$, (d,h) orientation $\theta(t)$ and (e,i) orbit $y_p(\theta)$. All lengths are made dimensionless using $L/2$, and time is normalized with the fiber velocity and channel length, i.e. $L/u_f$. Experimental trajectories correspond to the experiments shown on Fig. \ref{fig:glancing-pics} (d).}
		\label{fig:glancing-data}
		
	\end{sidewaysfigure}

\begin{figure}
\centering
		\includegraphics[width=\textwidth]{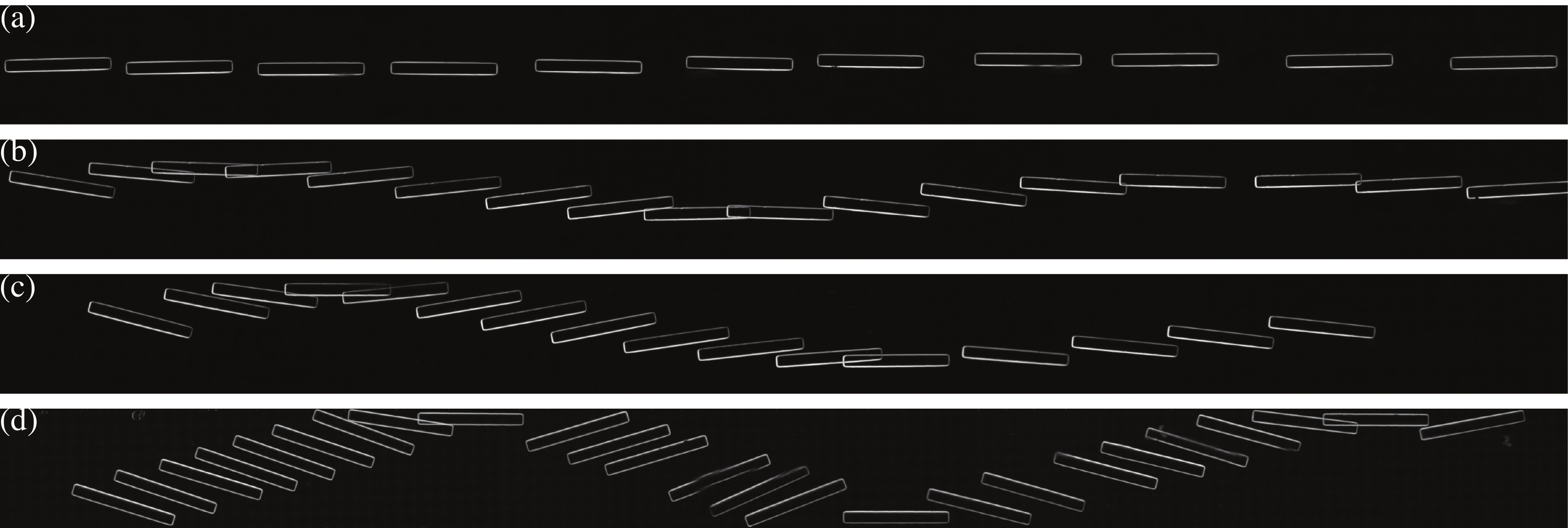}
		\caption{Experimental chronophotographies of fiber oscillations ($\beta=0.81$, $\xi=0.81$) for increasing initial angles: (a) $\theta_i=1.2^{\circ}$, (b) $\theta_i=9^{\circ}$, (c) $\theta_i=17^{\circ}$ and (d) $\theta_i=22^{\circ}$.}
		\label{fig:glancing-pics}
\end{figure}

We place the fiber at the center of the channel. When its initial orientation deviates from $\theta_i=0^{\circ}$, we find that the fiber exhibits oscillations that we report in Fig \ref{fig:illustration_russel}(b), Fig. \ref{fig:glancing-pics} and sketch in Fig. \ref{fig:glancing-data}(a); we call these oscillations \emph{glancing}. As the flow transports the fiber, the fiber drifts towards one of the side walls and rotates until parallel to the wall. Then, the angle of the fiber increases again and the fiber starts drifting away from the wall. The fiber thus oscillates between the two walls. These oscillations observed in experiments are recovered with our numerical model. The corresponding data, obtained experimentally and numerically, are given Fig. \ref{fig:glancing-data}.
	
We first note that the fiber keeps a nearly constant axial velocity, $u_f$, as it oscillates from one wall to the other (see the evolution of $x_p(t)$ in Fig. \ref{fig:glancing-data}(b,f)). This behaviour is recovered numerically. However, a detailed inspection of the data reveals that the velocity deviates around its average value (Inset in Fig. \ref{fig:glancing-data} (f)). Indeed, the fiber slightly accelerates and decelerates as it travels across the channel width. This small variation is within our experimental accuracy and can only be captured numerically.
From the evolution of $y_p(t)$ (Fig. \ref{fig:glancing-data}(c,g)), we observe that the fiber travels vertically with a drift velocity, $\dot{y_p}$, which too remains nearly constant throughout the channel width.

The trajectory can be described through the evolution of the fiber angle $\theta$ (Fig. \ref{fig:glancing-data}(d,h)). The fiber rotates when leaving/approaching a wall.
We thus observe that, as the fiber travels from the bottom wall ($y=-1$) to the top wall ($y=+1$), the angle first decreases then increases with $\theta<0^\circ$. Symmetrically, the angle first increases then decreases with $\theta>0^\circ$ when traveling from top to bottom wall. We can also note that the fiber rotates with a nearly constant velocity $\dot{\theta}$, with an inflection point around $\theta=0^\circ$ close to the walls.
The oscillations in the orientation $\theta$ and the position $y$ are shifted by half a phase, which is reminiscent of a mechanical pendulum for which momentum and restoring force are also out of phase.

Finally, we represent the path followed by the fiber with the orbit $y_p(\theta)$ (Fig. \ref{fig:glancing-data}(e,i)). We note that the fiber follows a closed orbit centered on (0,0) and bounded by $[-y_w, y_w]$ and $[-\theta_d$, $\theta_d]$, where we define $y_w$ as the position (dimensionless) at which the fiber is parallel to the wall, i.e. the minimum distance of approach of the wall is $|1-{y}_w|$, and $\theta_d$ as the maximal angle assumed by the fiber.


%

\begin{figure}
		\centering
		\includegraphics[width=\textwidth]{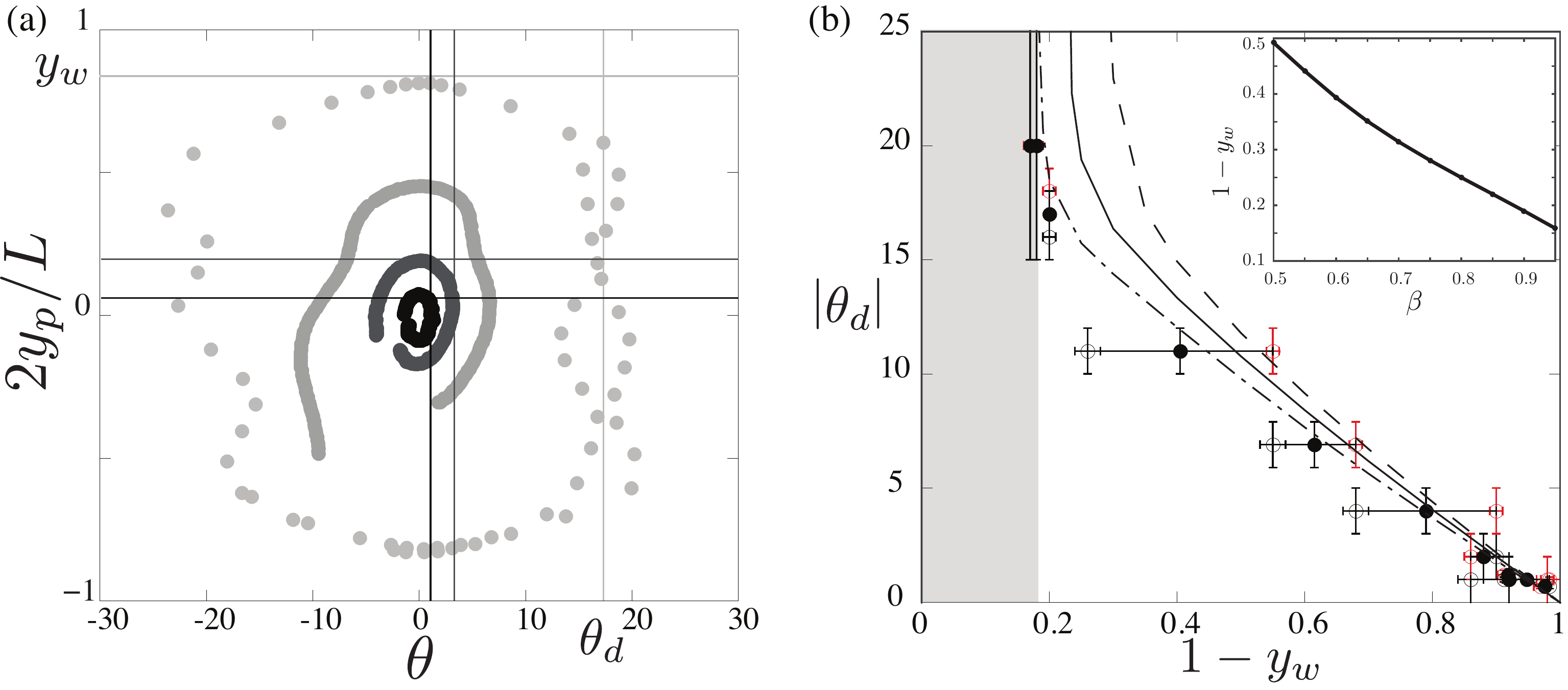}
		\caption{(a) Orbits corresponding to the chronophotographies presented in Fig. \ref{fig:glancing-pics}. (b)Maximum angle $\theta_d$ as a function of the distance to the wall $1-y_w$ for $\beta=0.81,\xi=0.81$. Angle as the fiber leaves the wall (open red circles), as it approaches the wall (open black circles). The full black circles correspond to the average values. Curves obtained numerically for $\xi = 0.8$ and $\beta = 0.7$ (dashed),$\beta= 0.8$ (full) and $\beta= 0.9$ (dash dotted line). Inset: Evolution of the minimal wall-fiber distance for $\theta_i=20^{\circ}$ as a function of the confinement $\beta$ for $\xi = 0.8$ and $\ell/h = 8$. }
		\label{fig:glancing-yw}
	\end{figure}

As the initial fiber angle increases, the amplitude of these oscillations increases, i.e. the fiber glances closer to the wall (Fig. \ref{fig:glancing-pics}). The corresponding orbits are presented in Fig. \ref{fig:glancing-yw}(a). For large angles, the orbit is oval, i.e. there is a region where the angle is constant (the fiber drifts without rotating near the center of the channel). For smaller angles $\theta_d$, i.e. small amplitude oscillations, the orbit is almost circular, i.e. the angle is nearly never constant. 
To each orientation $\theta_d$ corresponds a distance $y_w$ (and vice-versa). We plot the maximal angle $\theta_d$ as the a function of the minimal distance to the wall $|1-{y}_w|$ in Fig. \ref{fig:glancing-yw}(b) from experiments and numerical simulations. The numerical prediction agrees well with the experiments.
We observe that the fiber remains in the center of the channel (${y}_w\sim 0$) for angles $|\theta|\lesssim 2^{\circ}$, then the distance rapidly increases to reach ${y}_w\sim 0.8$ for $|\theta|\simeq 20^{\circ}$. We note that in this glancing regime the fiber does not touch the wall, i.e. there is always a finite distance between the fiber and the wall as highlighted by the shaded area in Fig. \ref{fig:glancing-yw}(b). The minimal wall-fiber distance is reached for $\theta_i\simeq 20^{\circ}$ for these conditions. We leverage on our numerical simulations to investigate the effect of the confinement $\beta$. The simulated trajectories indicate that the minimal wall-fiber distance decreases with increasing confinement (see inset in Fig. \ref{fig:glancing-yw}(b)). The amplitude of the oscillations thus increases with increasing confinements.

The trajectories presented in Figs. \ref{fig:glancing-pics} and \ref{fig:glancing-yw}(a) are not stable, i.e. the fiber leaves the wall with an angle slightly different than the initial approaching angle. This effect will be discussed further in \textsection \ref{sec:discussion}.


\subsubsection{Oscillations around $\theta=90^{\circ}$}
\begin{sidewaysfigure}
\vspace{5.15in}
		\centering
		\includegraphics[width=\textwidth]{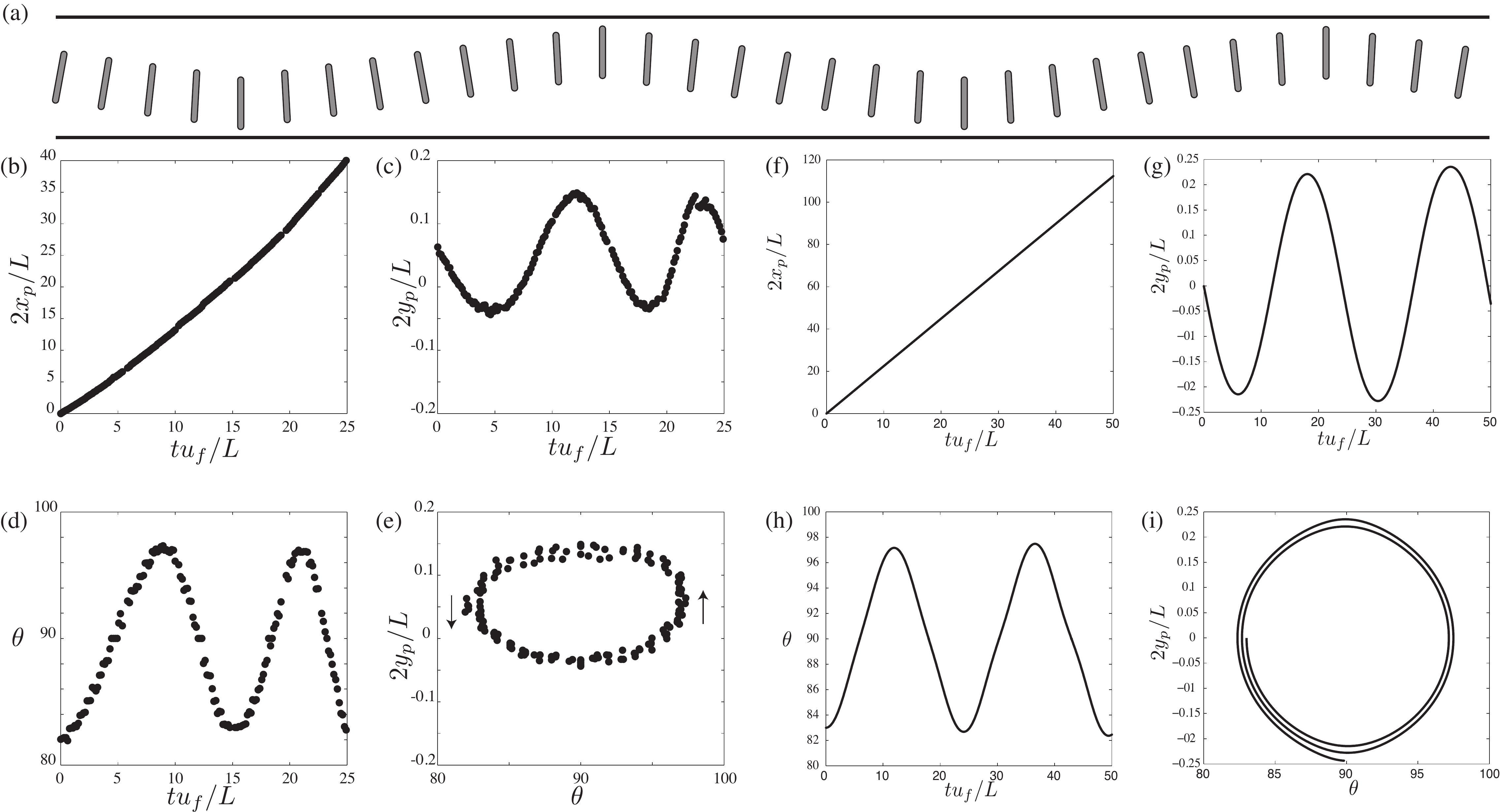}
		\caption{ (a) Sketch of the oscillation mode. (b-i) Trajectory of a fiber advected by a mean flow for $\beta=0.78$, $\xi=0.63$ and $\theta_i=82^{\circ}$ obtained experimentally (b-e) and numerically (f-i). (b,f) Axial position $x_p(t)$, (c,g) streamwise position $y_p(t)$, (d,h) orientation $\theta(t)$ and (e,i) Orbit $y_p(\theta)$. All lengths are made dimensionless using $L/2$, and time is normalized with the fiber velocity and channel length, i.e. $L/u_f$. The experimental trajectory corresponds to the experiments shown in Fig. \ref{fig:illustration_russel}.}
		\label{fig:reversing-data}
\end{sidewaysfigure}

When the initial angle is close to 90$^{\circ}$, the fiber exhibits qualitatively different oscillations as shown in Fig \ref{fig:illustration_russel}(b) and sketched in Fig. \ref{fig:reversing-data}(a). We call this second oscillation regime \emph{reversing}.  The corresponding data are shown in Fig. \ref{fig:reversing-data}(b-e). Again, our numerical model recover this oscillation regime. Similarly to the previous case, the fiber follows a closed orbit, this time centered around $\theta=90^{\circ}$. At the wall, the fiber is perpendicular to the flow direction, $\theta=90^{\circ}$. As the fiber travels from bottom to top wall, the angle first increases then decreases, contrary to what was found for glancing. Symmetrically, the angle first decreases then increases when travelling from top to bottom wall. This regime exists for a limited parameter range, and is in general difficult to observe. In this configuration, the fiber tips are close to the walls and the trajectory is sensitive to perturbations. On the experimental data presented in Fig.  \ref{fig:reversing-data}, we note a top/bottom asymmetry of the orbit due to small defects on the bottom channel wall, leading to a reduced distance $y_w$.

%

\subsubsection{Trajectories near the walls}
\label{sec:walllayer}

\begin{figure}
		\centering
		\includegraphics[width=\textwidth]{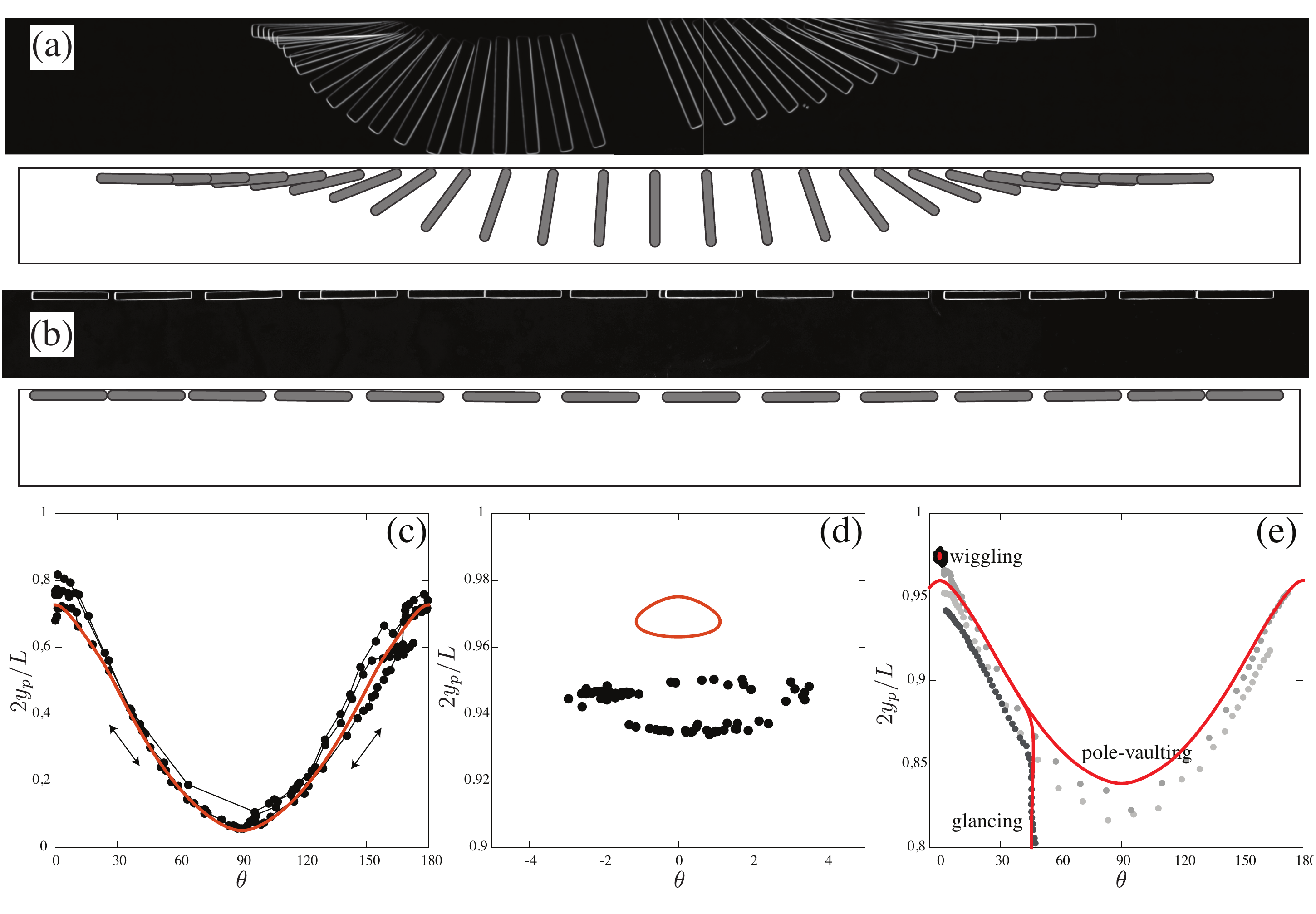}
		\caption{Chronophotographies of a fiber flowing near the wall for $\beta=0.81$, $\xi=0.81$, exhibiting (a) pole-vaulting and (b) wiggling, obtained both experimentally (top) and numerically (bottom). (c) Orbit for pole-vaulting for $\beta=0.78$, $\xi=0.93$. (d) Orbit for wiggling for $\beta=0.75$, $\xi=0.17$. (e) Trajectories near the walls for $\beta=0.86$, $\xi=0.16$. Numerical simulations with the same parameters are shown in red. For (e) the numerical pole vaulting and glancing trajectories become so close that they are indistinguishable near the wall.}
		\label{fig:pvwg}
	\end{figure}

As seen previously, there is a layer near the wall which is not accessible through glancing (shaded area in Fig. \ref{fig:glancing-yw}(b)). If the fibers are not initially placed at the center of the channel but in this layer, that is in the vicinity of the wall, we can access different types of trajectories. To investigate the occurrence of these trajectories we position the fiber horizontally ($\theta_i=0^\circ$) close to the wall $y_i \simeq y_w$ and observe the resulting trajectories as a function of the distance $1-y_i$ (Fig. \ref{fig:pvwg}).

For $y_i \leq y_w$, the fiber rotates and drifts away at a constant angle as described in the previous sections. As we decrease the distance to the wall $1-y_i$, another situation occurs, as presented in (Fig. \ref{fig:pvwg} (a)). The fiber remains near one wall, does not oscillate between the walls, but rotates around its tip in a \emph{pole-vaulting} motion \citep{Stover90}. In that case, the orbit is open (Fig. \ref{fig:pvwg} (c)). We note that only a small difference in distance to the wall leads to the transition between pole-vaulting and glancing motions (Fig. \ref{fig:pvwg} (e)). In fact, for large angle where fiber oscillation amplitudes are large, the fiber may alternate between those two types of motion as it moves downstream. Another situation occurs for even smaller fiber-wall distances $1-y_i$. The fiber remains in a layer near the wall, exhibiting small amplitude oscillations that we call \emph{wiggling} (Fig. \ref{fig:pvwg} (b)). The angle remains close to $0^{\circ}$ (Fig. \ref{fig:pvwg} (d)). The orbit is confined in a small parameter range (Fig. \ref{fig:pvwg} (e)). Our numerical simulations recover these two trajectories, as shown in Fig. \ref{fig:pvwg}. Numerically, we investigate the occurrence of these trajectories in a similar fashion as in the experiments: we place the fiber, either horizontally or vertically, at different distances of the wall and record its rotation rate $\dot{\theta}_p$ (Fig. \ref{fig:comp3d} (b)). For a fiber placed horizontally (i.e. parallel to the flow), we observe that the rotation rate becomes slightly positive very near the wall ($y_p\leq 0.9$), which indicates the transition to wiggling. Similarly, for a fiber placed vertically (i.e. perpendicular to the flow), the rotation rate becomes negative when approaching the wall, indicating a transition to pole-vaulting.

\subsection{State diagram}

We build a state diagram in the parameter space ($\theta$, $y_p$) to identify the various trajectories and isolate the regions in the parameter space where the different types of dynamics occur. We first present the diagram corresponding to a regime of high confinement, both transverse and lateral ($\beta=0.8$ and $\xi=0.8$). We report a remarkable agreement between experiments and simulations, showing that our 2D scheme captures the physics of this 3D problem. In addition, the numerical simulations give access to all possible trajectories to obtain a complete diagram. The experimental glancing orbit shows a spiraling behavior that is absent in the simulation, and that will be discussed further in \textsection \ref{sec:discussion}.
\begin{figure}
		\centering
		\includegraphics[width=.8\textwidth]{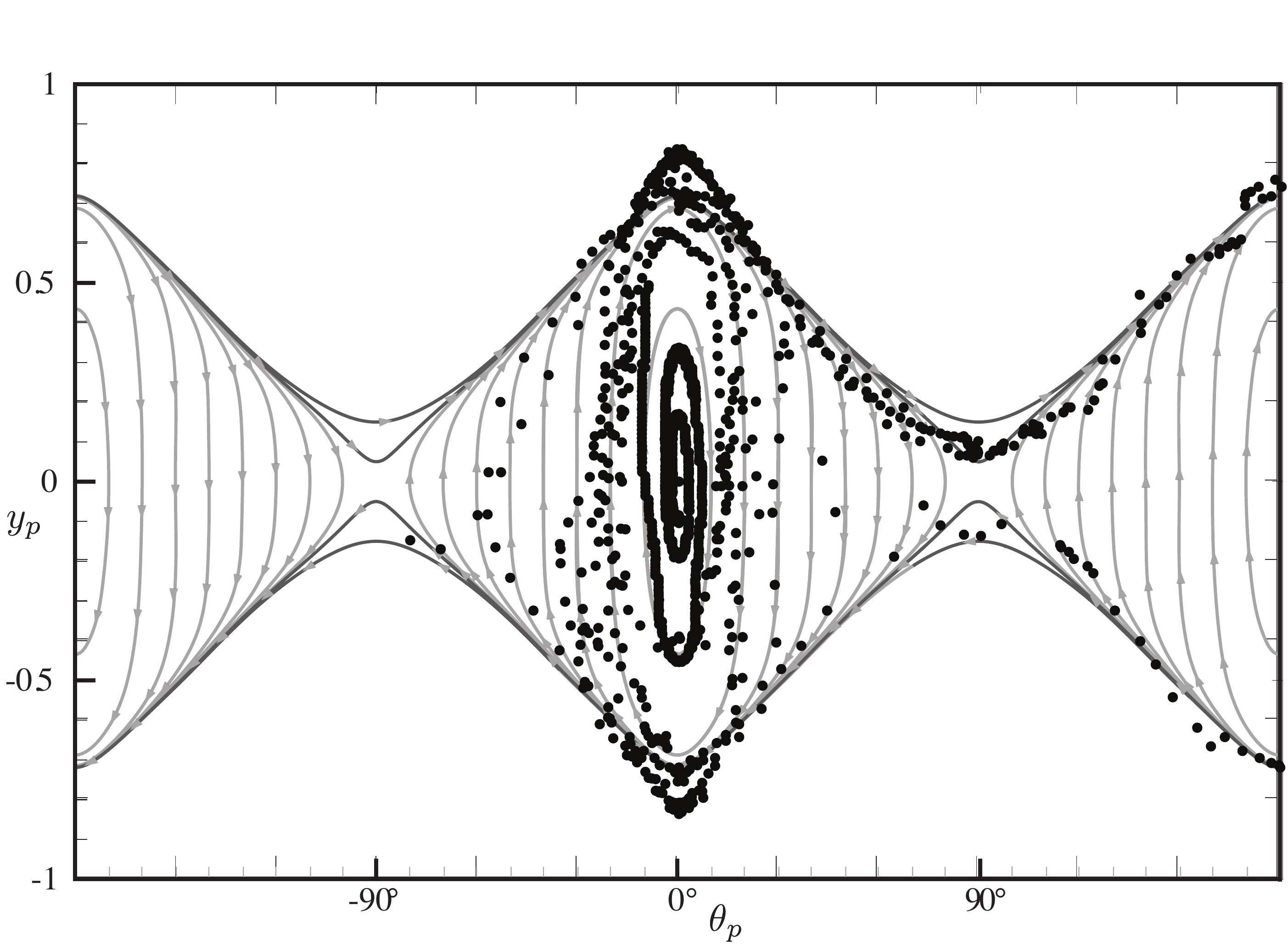}  
		\caption{Experimental ($\beta\simeq 0.8$, $\xi\simeq0.8$, black dots) and numerical state diagram ($\beta = 0.8$, $\xi = 0.8$, gray lines). Experimental results retrace the numerical orbits for glancing (around $\pm 30^\circ$) in light gray and pole vaulting in dark gray. The region in white corresponds to impossible configurations, i.e. the fiber tip touches the wall. }
		\label{fig:diagexp}
	\end{figure}
	
We can observe the pole-vaulting orbits, centered on $\theta=90^{\circ}$, and the glancing orbits, i.e. oscillations around the fixed point at $\theta=0^{\circ}$. For these values of $\beta$ and $\xi$, reversing oscillations are never observed, neither experimentally nor numerically.

The obtained state diagram is reminiscent of that of an undamped perfect pendulum \citep{Strogatz}.  It is characteristic of an Hamiltonian system, with time-reversal symmetry. In the present case, this property does not result from the absence of dissipation in the system, but from the symmetries of the fiber and the reciprocal properties of the Stokes equations. The orbits can be categorized exactly like for a pendulum. The pole-vaulting orbits are free (unbounded) trajectories, while the glancing orbits are bound states and both are separated by a separatrix. The state diagram is organized around two centers, the stable $\theta=0^\circ, y=0$ horizontal position and the structurally unstable hyperbolic center corresponding to the vertically aligned fiber $\theta= 90^\circ, y=0$. 

\begin{figure}
		\centering
		\includegraphics[width=0.8\textwidth]{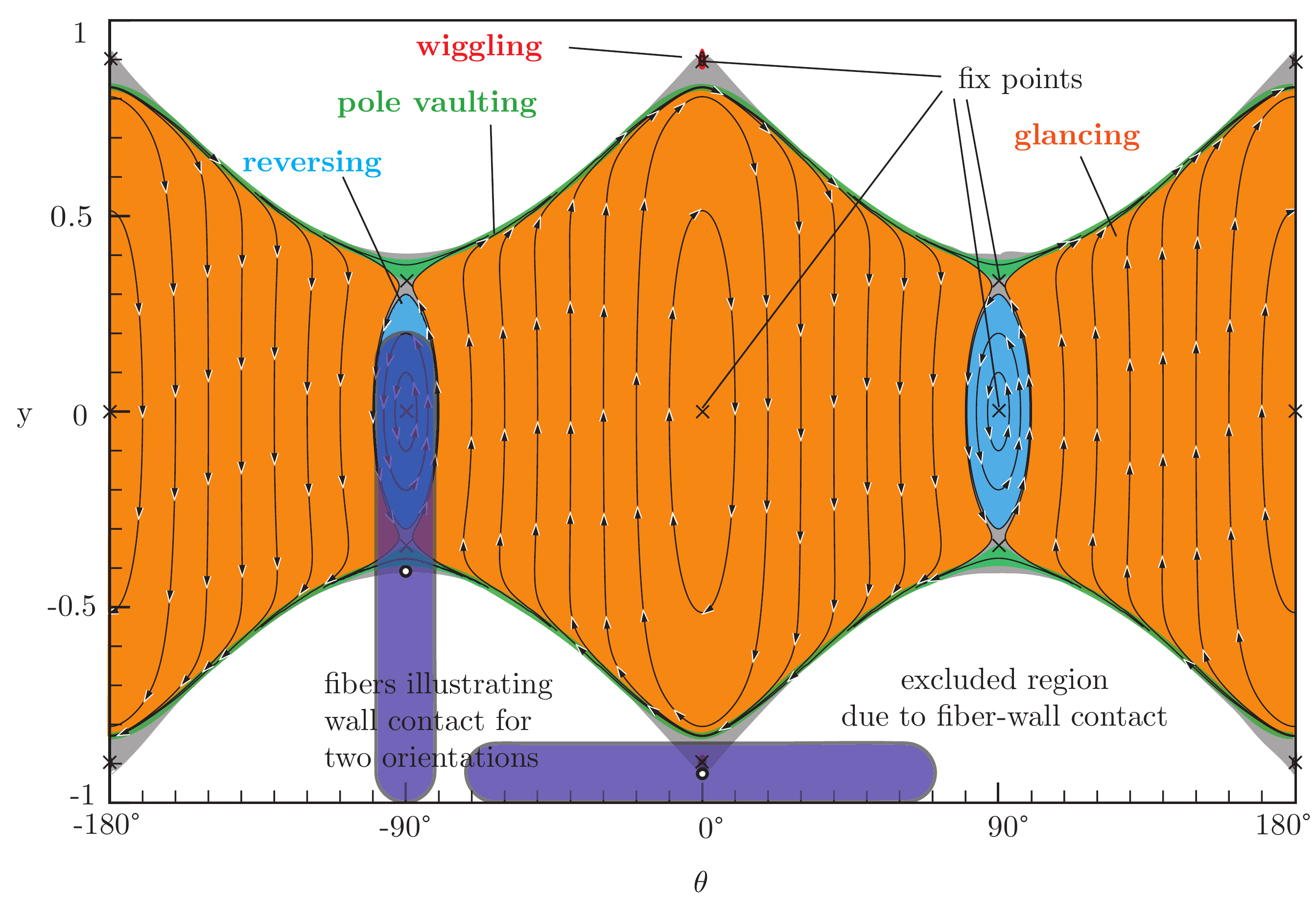}
		\caption{Numerical state diagram for $\beta=0.8$, $\xi=0.6$. displaying all four orbit types: glancing, reversing, pole vaulting and wiggling together with their associated fix points. Separatrices exist in between the colored regions, where the numerical scheme was unable to resolve trajectories the color was left grey. The region in white corresponds to impossible configurations, i.e. the fiber tip touches the wall.}
		\label{fig:diagnum}
	\end{figure}

While, as stated earlier, reversing orbits could not been observed for the parameters of Figure \ref{fig:reversing-data}, they can be obtained for lower values of the lateral confinement $\xi$. For $\beta=0.8$, reversing is only obtained for $\xi \leq 0.6$ as observed in the experiments (Fig. \ref{fig:reversing-data}). This regime is further explored numerically and a complete state diagram for this value is presented in Fig. \ref{fig:diagnum}.  In addition to glancing, pole-vaulting and wiggling, reversing can be observed in the vicinity of $\theta= 90^\circ$. This diagram is also characteristic of Hamiltonian dynamics, but it is more complex than that of a simple pendulum. It has stable centers in $\theta=0^\circ, 180^\circ, y=0$, $\theta= \pm 90^\circ, y=0$ and $\theta= 0^\circ,180^\circ, y \approx \pm 0.92$ as well as unstable hyperbolic centers in $\theta= \pm 90^\circ, y \approx\pm 0.32$ and $\theta= 0^\circ, 180^\circ, y\approx\pm 0.9$. These fix points structure the phase space which has dynamics with three types of bound states, glancing, reversing and wiggling orbits and separatrices that connect the hyperbolic points.

From a theoretical point of view, the motion of a fiber is completely described by the coordinates $y_p$ and $\theta_p$ for a given configuration in a rectangular channel . Therefore, in theory, the trajectories in the state diagram are, apart from singular points, free of intersections and consequently closed loops. However, in both experimental and numerical diagrams, the orbits are close to each other, especially near the walls (along the pole-vaulting separatrix) and at the glancing-reversing limit. As a result, the fiber easily jumps from one orbit to another, as we will discuss in the next paragraph.

\section{Discussion}
\label{sec:discussion}

\subsection{Glancing}

\begin{figure}
		\centering
		\includegraphics[width=0.9\textwidth]{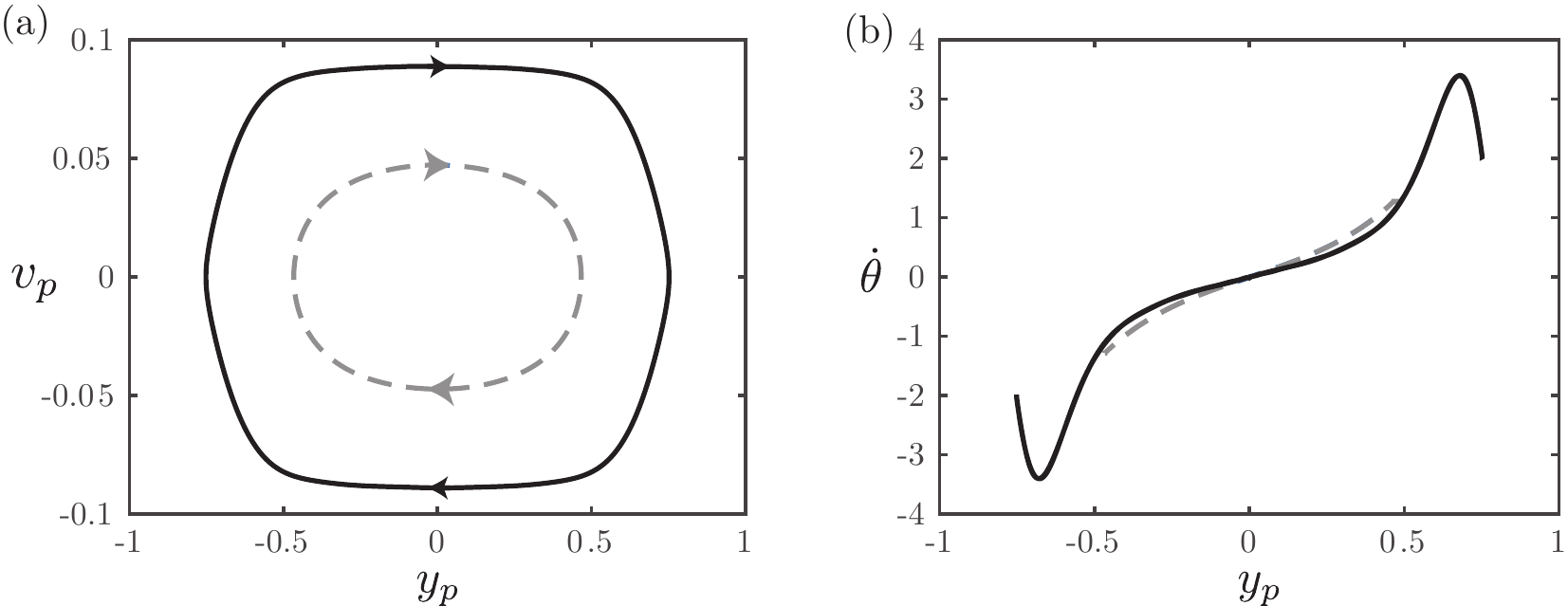}
		\caption{Evolution of the drift velocity $v_p=\mathrm{d}y_p/\mathrm{d}t$ (a) and the rotation velocity $\dot{\theta}=\mathrm{d}\theta/\mathrm{d}t$ (b) as a function of the fiber position $y_p$ as the fiber oscillates between the walls. Computed for $\beta=0.8$ and $\xi=0.8$ and for two initial angles $\theta_i=10^{\circ}$ (dashed line, gray) and $\theta_i=20^{\circ}$ (full line, black). }
		\label{fig:velocitiesdiscussion}
	\end{figure}

We first investigate the glancing trajectories. The trajectory of a fiber is a combination of drift, due to drag anisotropy, and rotation due to the presence of the lateral walls. We can compute numerically both the vertical drift velocity $\dot{y_p}$, and the rotation velocity $\dot{\theta}$. We compare the evolution of these velocities as the fiber oscillates between the walls for two different initial angles (Fig. \ref{fig:velocitiesdiscussion}). As the fiber travels from the bottom wall to the center of the channel, the drift velocity increases (with $\dot{y_p}>0$) while the fiber rotates away for a horizontal orientation ($\dot{\theta}<0^\circ$). Near the center of the channel, the rotation velocity $\dot{\theta}=0^\circ$ and the curve presents an inflection point; the orientation of the fiber is thus constant, which leads to a constant drift velocity as seen in Fig. \ref{fig:velocitiesdiscussion}(a). As the fiber travels away from the center and approaches the wall, the drift velocity rapidly decreases to reach zero when the fiber is horizontal, and the rotation velocity increases as the fiber reorients ($\dot{\theta}>0^\circ$). The top to bottom trajectory is symmetrical (with $\dot{y_p}<0$).

We note that the drift velocity strongly depends on the fiber angle ($\dot{y_p}(\theta=20^\circ)\simeq 2~ \dot{y_p}(\theta=10^\circ)$). Indeed, the drift velocity, given in eq.\eqref{eq:udrift}, is proportional to $\cos\theta\sin\theta$. The drift velocity thus increases almost linearly with the orientation angle up to $\theta=45^{\circ}$. 
On the contrary, the rotation velocity is nearly independent of the fiber angle.

These results qualitatively explain the observations made on the fiber trajectories. The fiber drifts and rotates simultaneously. Since the rotation velocity is nearly constant, the maximum displacement $|y_p|$ of the motion depends on the drift velocity: the fastest the drift velocity, the further the fiber can travel in the channel. Indeed, as we increase the initial angle, the rotation velocity is unaffected while the drift velocity strongly increases, leading to larger amplitude oscillations (i.e. the fiber travels a longer distance before it is rotated, hence glances closer to the wall). On the contrary, small angles lead to slow drift velocities and so to small amplitude oscillations, so that a fiber whose orientation is close to 0 remains in the center of the channel.

Furthermore, as we increase the confinement, we can tune the difference between $u_{\parallel}$ and $u_{\perp}$, and thus increase the drift velocity ($\dot{y_p}\propto (u_{\parallel}-u_{\perp})$) and the oscillation amplitude, as observed experimentally and numerically.

\subsection{Transition between glancing and reversing}

\begin{figure}
		\centering
		\includegraphics[width=\textwidth]{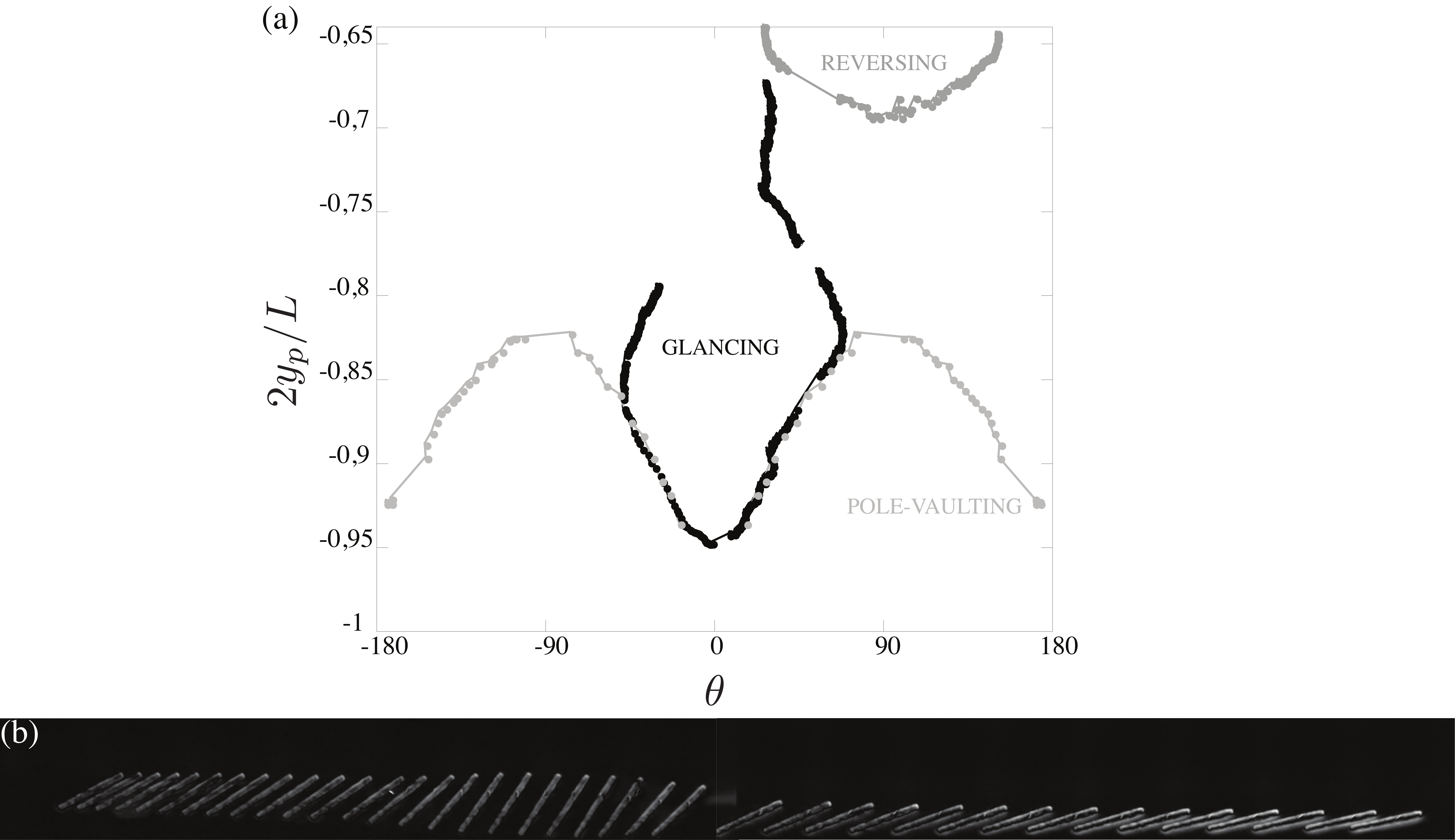}
		\caption{Trajectory obtained experimentally for $\beta=0.76$, $\xi=0.14$. (a) Orbits (b) Chronophotography of the limit trajectory between glancing, reversing and pole-vaulting. The channel being wide, we only display a region near the bottom wall, i.e. $-0.5 > y_p > -1$.}
		\label{fig:revglanc}
	\end{figure}

During an oscillation in the reversing regime, the evolution of the drift velocity is analogous to glancing. Similarly, the drift velocity decreases when approaching $\theta=90^{\circ}$ so that the amplitude of the oscillations decreases as the fiber angle gets closer to $\theta=90^{\circ}$ (as can be seen from the state diagram Fig. \ref{fig:diagnum}). However, the sign of the rotation speed is inverted (i.e. the angle decreases as the fiber leaves the wall and increases as the fiber approaches a wall, in an opposite fashion compared to the glancing regime). The transition between glancing and reversing can be obtained by looking at the sign of the rotation speed.
We obtain the critical angle $\theta^{\star}$ for which the transition occurs for different confinements. For $\xi\sim0.8$, the value of $\theta^{\star}$ remains close to $90^{\circ}$, and slightly decreases for increasing confinement $\beta$. This explains why reversing is difficult to observe at high lateral confinement, where the region of reversing in the parameter space is narrow around $90^{\circ}$.
The reversing region expands when decreasing $\xi$. Experimentally, we indeed observe systematically reversing in the wide channels ($\xi=O(10^{-2})$). However, in such wide channels it is not possible to follow an entire oscillation as the fiber travels through a long distance between the walls. We thus limit ourselves to the trajectories near one wall (Fig. \ref{fig:revglanc}). Numerically, we observe that the reversing region expands when decreasing $\xi$, i.e. $\theta^{\star}$ decreases to approach 0. While in highly confined channels ($\xi\sim1$), reversing is confined in a narrow region around $90^{\circ}$, for low confinement this region can reach larger regions (as observed in Figs. \ref{fig:diagnum} and \ref{fig:revglanc}).
We also note that the limit orbit between reversing and glancing is peculiar (see the line separating the yellow and blue regions in Fig. \ref{fig:diagnum}); indeed, the angle first increases then decreases as the fiber approaches the wall. This can also be observed experimentally (Fig. \ref{fig:revglanc}). 
The corresponding orbits are given for $\xi = 0.14$ and initial angles around $25^{\circ}$, at the limit of glancing, reversing and pole-vaulting. We note the reversing (dark gray) and pole-vaulting (light gray) trajectories. For the glancing trajectory (black), the fiber drifts near the center of the channel, its angle increases as in the reversing trajectory, however the angle then decreases to reach zero in a glancing motion. Similarly, upon leaving the wall, the angle first increases then decreases.
	

\subsection{Transition between wiggling, pole-vaulting and glancing}

While the flow within the rectangular channel is a plug flow throughout most of the width of the channel, there is a small layer near the walls, of size $\sim H$, where there is shear due to the no-slip condition at the wall. For the data presented in Fig. \ref{fig:glancing-yw}(b), the boundary layer thickness corresponds to $|1-y_w|\simeq 0.1$, i.e. during glancing the fiber always remains outside the boundary layer. When placed within the boundary layer, the fiber exhibits different trajectories, i.e. either a pole-vaulting or a wiggling motion.

\begin{figure}
		\centering
		\includegraphics[width=0.49\textwidth]{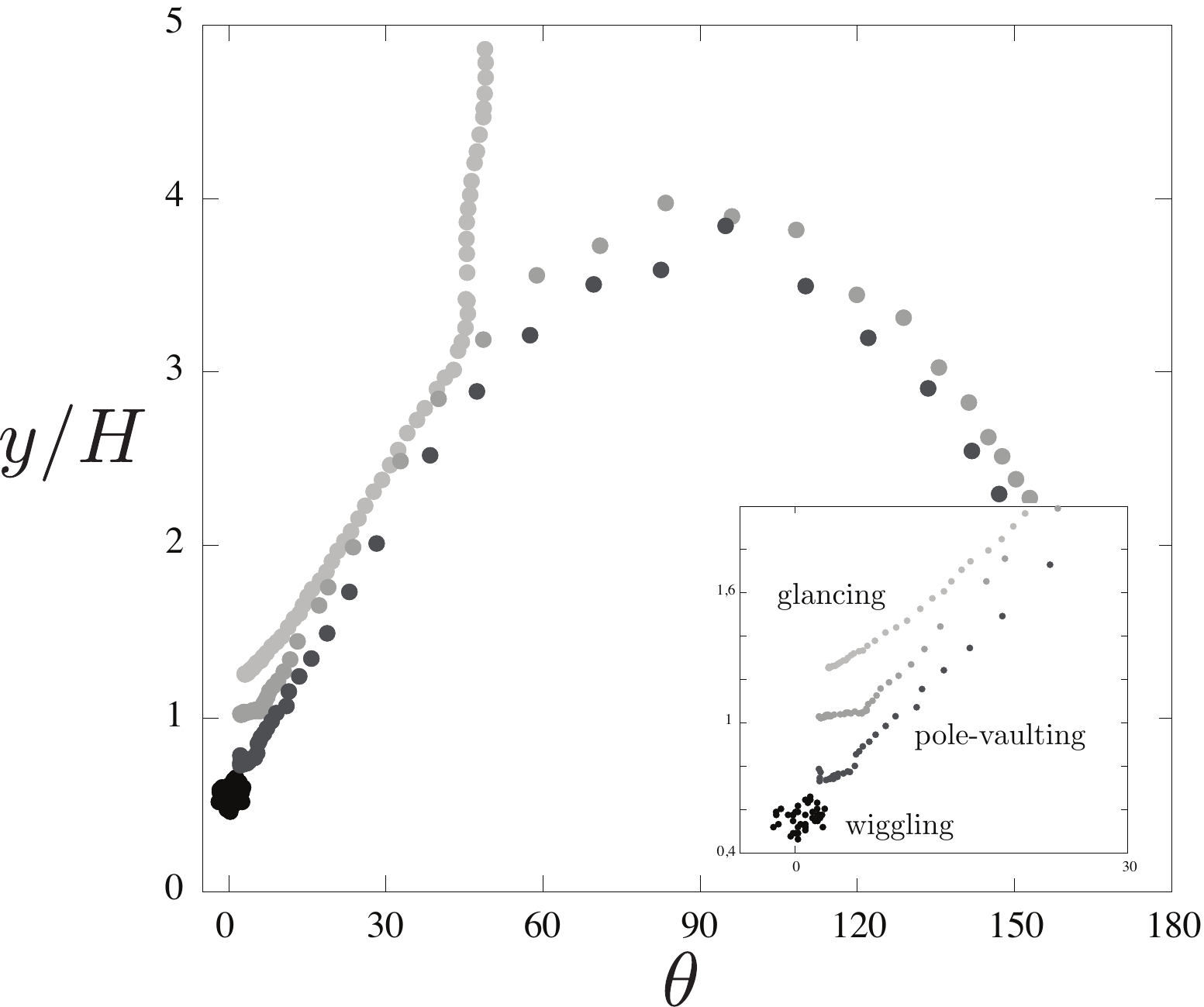}
			\caption{Trajectories near the wall (same data as Fig. \ref{fig:pvwg} (e)) rescaled with channel height $H$. The inset is a zoom of the data around $y/H=1$.}
		\label{fig:nearwalls2}
	\end{figure}

We can discriminate between the different trajectories using the boundary layer thickness. We rescale the data corresponding to the different trajectories near the walls (Fig. \ref{fig:pvwg} (e)) with the boundary layer thickness $H$, as shown in figure \ref{fig:nearwalls2}. We indeed observe that wiggling exists for $y<H$, pole-vaulting for $y\sim H$, and glancing for $y> H$. As the fiber is entirely within the boundary  layer ($y<H$), it oscillates without leaving the wall in a wiggling motion. If the fiber tip is within the boundary layer $y\sim H$, the fiber may be rotated around its tip, thus exhibiting a pole-vaulting motion. For larger distances, the fiber is not affected by the shear and simply rotates away from the wall in a glancing motion.

	\color{black}

\subsection{Transition between reversing and pole vaulting}
\begin{figure}
		\centering
		\includegraphics[width=\textwidth]{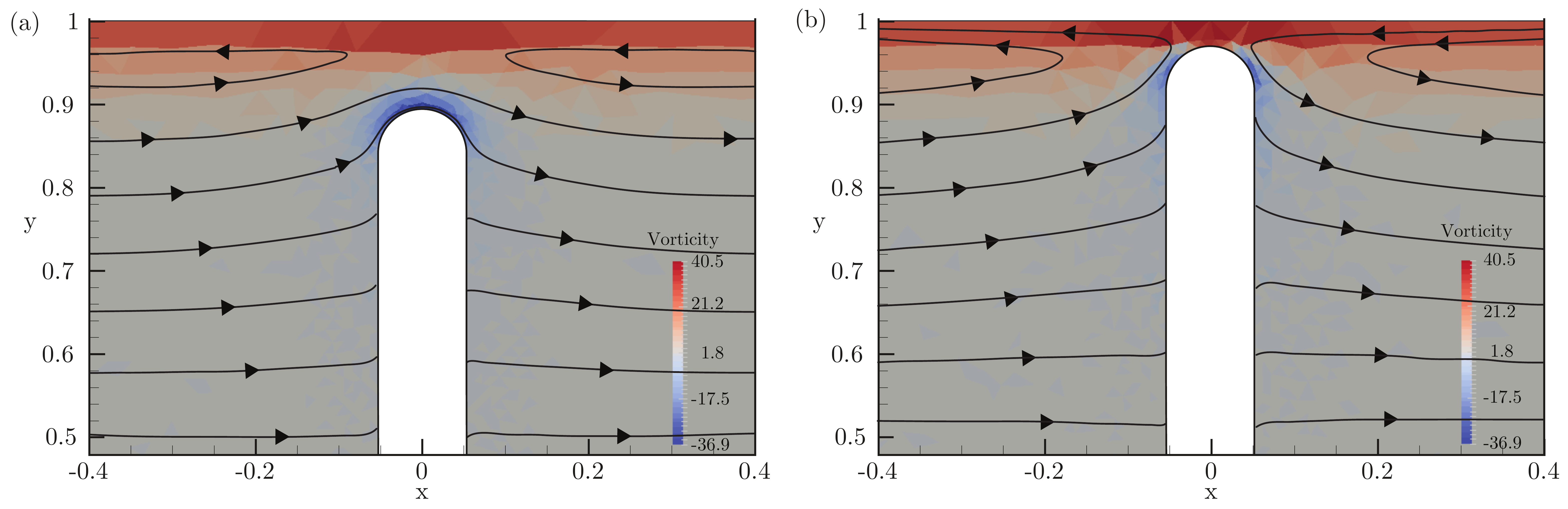}
		\caption{3D Simulation of a moving fiber with $\beta = 0.8$ in a wide channel $\xi = 0.5$. Shown are the data from the center plane in $z$.    Streamlines are drawn with respect to the fibers mean velocity for a) reversing fiber and b) pole vaulting fiber.}
		\label{fig:ecoulementRPV}
	\end{figure}

In a sufficiently wide channel, i.e. with $\xi$ lower than $0.8$ in the case $\beta = 0.8$ and $\ell/h = 8$, a fiber can show reversing or pole vaulting when oriented perpendicular to the flow. For the description of reversing the presence of a potential Hele-Shaw flow (corresponding to a plug flow) is sufficient. A reversing motion has been indeed justified by image potentials \citep{Leman14,Uspal2013}. As the fiber enters the shear boundary layer near the lateral walls, potential flow is no longer valid and one observes a competition between a tip acceleration due to the leaking flow and tip deceleration due to wall friction. In figure \ref{fig:ecoulementRPV} (a) one sees a fiber performing a reversing motion, where the liquid flows through the gap between fiber tip and lateral wall. A negative vorticity is observed at the fiber tip. In comparison, if the fiber is too close to the wall no fluid leakage is observed between the fiber tip and the wall. A positive vorticity close to the fiber tip leads to a pole vaulting motion (Fig. \ref{fig:ecoulementRPV} (b)).

\subsection{Perturbations and damped oscillations}
\begin{figure}
		\centering
		\includegraphics[width=0.8\textwidth]{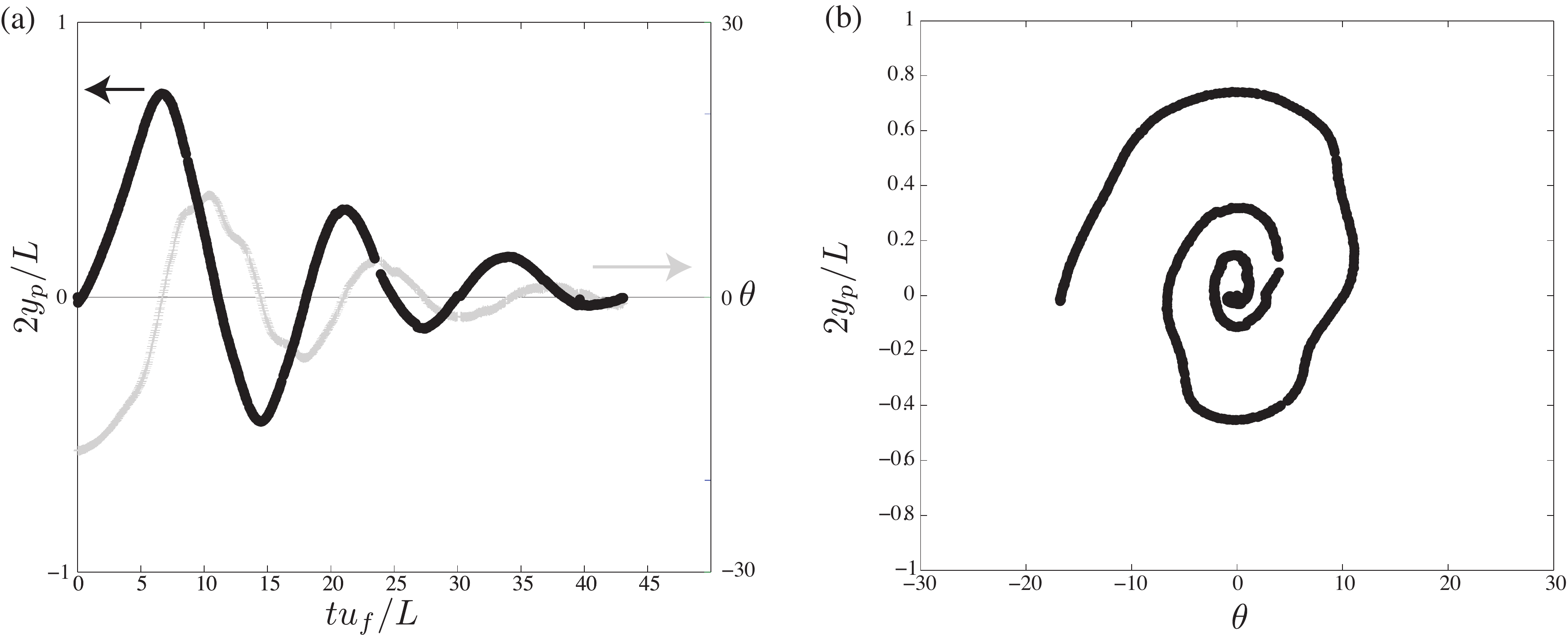}
		\caption{Damped oscillations obtained experimentally for $\beta=0.8$, $\xi=0.8$. (a) Position $y$ and angle $\theta$ as a function of time. (b) Spiralling orbit. }
		\label{fig:damping}
	\end{figure}

The system is sensitive to perturbations. Any perturbation leads to a small change in orientation, thus a change in trajectory. In particular, we often observe in experiments that the fiber leaves the wall with an angle slightly smaller than the initial approaching angle. The amplitude of oscillations may thus decrease as the fiber moves along the channel. This effect may be important, leading to damped oscillations as presented in Fig. \ref{fig:damping}. The amplitude of the oscillations decreases and the fiber moves towards the center of the channel, until reaching the equilibrium position, i.e. the fix point ($\theta=0^\circ$, $y_p=0$), staying parallel to the flow in the center of the channel $y_p=0$. \cite{Uspal2013} showed that a fore-aft asymmetry also leads to the reorientation of the particle towards $\theta=0^\circ$. The analogy with the pendulum's trajectories evoked earlier (\textsection 3.3) can be pushed further. The addition of a slight imperfection, here taking the form of a small experimental spatial asymmetry or an experimental or numerical time-symmetry breaking, acts in a similar way as the addition of a small damping in the pendulum \citep{Strogatz}. While the unstable fix points disappear, the previously neutrally stable centers become focal points attracting the dynamics through spiraling trajectories. This behavior results from the structural instability of the hyperbolic fix points.

\section{Conclusion}

We have investigated experimentally and numerically the motion of rigid fibers, confined in a Hele-Shaw cell, and transported by a pressure-driven flow. Due to the transverse and lateral confinements, complex fiber oscillations between the channel walls are observed in experiments, as well as pole-vaulting and 'wiggling' when the fiber is placed near the walls. We have shown that the trajectories can be controlled by adjusting the geometry of the fiber and the channel. We have in particular shown the effect of the transverse confinement (height of the channel), which controls the magnitude of the drag anisotropy, thus the drift angles and as a consequence the amplitude of the oscillations.

We have developed a model using the $2D$~Brinkman equations that correctly describes the dynamics. By numerically solving the model equations, we can quantitatively reproduce and predict the experimental trajectories, as well as qualitatively explain the different motions and transitions. Due to the ideal setting of a numerical simulation with perfectly smooth walls and rigid fibers, clear distinctions can be made between stable and unstable flow configurations in the absence of fabrication defaults.

The glancing and reversing motion of the fiber are controlled by the pressure field that develops around the fiber. As a consequence, these can be, at least qualitatively \citep{Uspal2013}, described by the Darcy equation. Pole-vaulting and wiggling motion is driven by shear flow close to the walls and can therefore not be described by the Darcy equation since there is no in-plane shear contribution, requiring therefore a more elaborate approach. Interestingly, a qualitative analysis of sedimenting flows can rely solely on the Darcy equation since the pole-vaulting motion can be thought of as being part of the reversing motion and the wiggling motion being part of the glancing motion. Only in pressure driven flows, when the drift direction is about perpendicular to the fiber orientation, pole vaulting and wiggling exist as singular limits in a boundary layer close to the walls.

From looking at the state diagram (Fig. \ref{fig:diagnum}) one sees that the fiber trajectories come very close in certain regions, which gives rise to random changes in orbits due to small perturbations, which we observed experimentally as well as numerically. The limited numerical predictability thus reflects a physically unpredictable state.

 An asymmetrically designed particle has been observed to create a driving force that keeps the perturbed particle in a well defined orbit \citep{Uspal2013}, where regions of close contact of orbits from different oscillatory regimes are avoided.
Within the framework of our numerical method it is possible to address the behavior of tailored asymmetric particles, as well as multiple particle interaction. In addition, we correctly account for the effect of the confinement, which can be used to tune the strength of the drag anisotropy, thus the trajectories of the particles. We believe that this extension will be useful for the investigation of design principles for the transport of rigid objects in microfluidic devices.

Our efforts in developing and implementing the routines presented in this work are made available for the community. We share the source codes and documentation of the simulation tool (\url{ulambator.sourceforge.net}) and provide online tutorials and examples (\url{lfmi.epfl.ch/ulambator}).


\subsection{Acknowledgements}

The authors acknowledge the help of Caroline Frot with microfabrication, and useful discussions with Howard A. Stone. The European Research Council is acknowledged for funding the work through a starting grant (ERC SimCoMiCs 280117). Agence Nationale de la Recherche (ANR-DefHy) is acknowledged for partial funding of this work and travel support.

\appendix

\section{3D Stokes solution with the Finite Element Method}
\label{ap:fem}
The FEM simulation of the 3D Stokes equation is done with FreeFEM++ \citep{Hecht2012}, where the mesh is generated in a layered structure. Owing to the symmetry in the $z-$direction only half of the domain is solved. Due to the sparse matrices, the force free condition on the fiber cannot be imposed directly. Instead, four independent problems are solved in series, with four different boundary conditions: outer flow, fiber moving at $U_p=1$, fiber moving at $V_p=1$ and fiber rotating at $\dot \theta_p=1$. The torque and the $x-$ and $y-$ forces on the fiber are saved, and the fiber displacement is obtained from a force balance once the four calculations are completed.
We use $P_1^b/P_1$ elements, well suited for geometries with sharp edges, where the additional degree of freedom for the velocity description (the so-called bubble) avoids spurious pressure modes. The mesh size is about $1.3$ million nodes, which results in about $5$ millions unknowns. The convergence is given for one example in the inset in figure \ref{fig:comp3d}.



\bibliographystyle{jfm}

\end{document}